\documentclass[11pt]{article}
\usepackage{latexsym} \usepackage{epsf}
\usepackage{a4}
\usepackage{amsfonts}
\renewcommand{\theequation}{\thesection.\arabic{equation}}

\makeatletter

\expandafter\let\expandafter\reset@font\csname reset@font\endcsname

\makeatother

\def\be{\begin{equation}}
\def\ee{\end{equation}}
\def\bea{\begin{eqnarray}}
\def\eea{\end{eqnarray}}
\def\dd{\partial}
\def\'{\prime}
\def\a{\alpha}
\def\b{\beta}

\def\ba{\begin{array}}
\def\ea{\end{array}}
\def\sh{{\rm{sh}}}
\def\ch{{\rm{ch}}}
\makeatletter
\def\binrel@#1{\begingroup
  \setboxz@h{\thinmuskip0mu
    \medmuskip\m@ne mu\thickmuskip\@ne mu
    \setbox\tw@\hbox{$#1\m@th$}\kern-\wd\tw@
    ${}#1{}\m@th$}%
  \edef\@tempa{\endgroup\let\noexpand\binrel@@
    \ifdim\wdz@<\z@ \mathbin
    \else\ifdim\wdz@>\z@ \mathrel
    \else \relax\fi\fi}%
  \@tempa
}
\let\binrel@@\relax
\def\overset#1#2{\binrel@{#2}%
  \binrel@@{\mathop{\kern\z@#2}\limits^{#1}}}
\def\underset#1#2{\binrel@{#2}%
  \binrel@@{\mathop{\kern\z@#2}\limits_{#1}}}
\makeatother
\newcommand{\p}{^{\prime}}
\newfont{\bbd}{msbm10 scaled\magstep1}

\begin{document}
\hfill LU-ITP 2003/23

{\begin{center}
{\LARGE {Universal R operator \\
with Jordanian deformation of conformal symmetry} } \\ [8mm]
{\large  S. Derkachov$^{a}$,
D. Karakhanyan$^b$\footnote{e-mail: karakhan@lx2.yerphi.am} \&
R. Kirschner$^c$\footnote{e-mail:Roland.Kirschner@itp.uni-leipzig.de} \\
[3mm] }
\end{center}

\begin{itemize}
\item[$^a$]
Department of Mathematics, St Petersburg Technology Institute, \\
Sankt Petersburg, Russia
\item[$^b$]
Yerevan Physics Institute , \\
Br.Alikhanian st.2, 375036, Yerevan , Armenia.
\item[$^c$] Naturwissenschaftlich-Theoretisches Zentrum und
Institut f\"{u}r Theoretische Physik, Universit\"{a}t Leipzig, \\
Augustusplatz 10, D-04109 Leipzig, Germany
\end{itemize}

\vspace{3cm}

\begin{center}
{\bf Abstract}
\end{center}
\noindent
The Jordanian deformation of $sl(2)$ bi-algebra structure is studied in view
of physical applications to breaking of conformal symmetry in the high
energy asymptotics of scattering. Representations are formulated  in terms
of polynomials, generators in terms of differential operators. The deformed
$R$ operator with generic representations is analyzed in spectral and integral
forms.


\renewcommand{\refname}{References.}
\renewcommand{\thefootnote}{\arabic{footnote}}
\setcounter{footnote}{0}
\setcounter{equation}{0}

\renewcommand{\theequation}{\thesection.\arabic{equation}}

\section{Introduction}
\setcounter{equation}{0}

In the last decade high-enegy scattering in gauge field theories has
become  a new area of application of integrable quantum systems
\cite{LevPadua,FK}. In several cases the effective interaction
appearing in the Regge and the Bjorken asymptotics of scattering is
determined by Hamiltonians of integrable chains, where the
representations on the sites are infinite-dimensional. In a number of
papers  the methods of integrable systems have been reformulated and
developed for the special needs of these new applications, e.g. 
\cite{DeVega:2001pu, Derkachov:2001yn}.
They have been applied to the renormalization of higher twist operators
\cite{Braun:1998id,Belitsky:1999qh}.
Integrable structures have been encountered also in recent studies of
composite operators in ${\mathcal N} $ = 4 super Yang-Mills theory in the
large $N_C$ limit motivated by questions related to the AdS/CFT
hypothesis \cite{Beisert:2003yb,Dolan:2003uh} .

In the context of integrable chains with non-compact representations on
the sites it is convenient to consider wave functions describing the
states on the sites. In the case of the Bjorken asymptotics the one-site
wave function depends on one variable, the position on the light ray.
Conformal transformation of the light ray is the relevant symmetry
here, the underlying {\sl Lie} algebra is $sl(2)$. Infinitesimal
symmetry transformation of the wave functions act as differential
operators.

Owing to these applications solutions of the quantum Yang-Baxter
equation (QYBE) have been studied with generic representations of the
symmetry algebras $sl(2)$, $sl(2|1)$ and the $q$-deformation of $sl(2)$
\cite{Derkachov:2000ne,Derkachov:2001sx,Karakhanyan:2001wr}.
 A scheme has been developed relying on known methods
\cite{KS80,TTF,Nankai,LesHouches} and 
resulting in formulations suitable for the mentioned applications.

In the present paper we treat along these lines the non-standard or
Jordanian ($\xi$) deformation of $sl(2)$. Whereas there is an extensive
literature on the standard $q$ deformation of Lie bi-algebra structure
and on its application, the case of $\xi$- deformation did not attract
comparable attention so far, although the related literature is
numerous and cannot be review here. Some early papers concerning this
subject are \cite{Manin, Zakrewski,Stolin,Wess,
Kupershmidt,Ohn,Gerstenhaber,Ogievetsky}.

The concepts of Yangians and of their deformation by twist have been
developed by Drinfeld \cite{DrinfeldT,DrinfeldY}.
The basis commutation relation of $\xi$ deformed $sl(2)$ algebra have
been written first by Ohn \cite{Ohn}.
The explicite form of the twist transformation appeared first in
\cite {Ogievetsky, Gerstenhaber}. The twist construction of the deformed
Yangian has been described e.g.in \cite{KS,KhST}.
The structure of $\xi$- deformed representations has been studied e.g. in
\cite{Abdesselam,Jeugt}.
Applications to spin chains have been discussed in \cite{KS}
and to the modification of space-time structure  e.g. in
\cite{Vladimirov,Ballesteros,Lukierski}.

From the viewpoint of conformal symmetry representations the $\xi$
deformation of $sl(2)$ leaves the generator of translation undisturbed
but deformes the one of dilatation. This means momentum conservation
holds, whereas scale symmetry is broken. By the deformation parameter a
(length or momentum) scale is introduced. It remains to be investigated
whether the pattern of scale symmetry breaking by $\xi$ deformation,
which will be worked out to some detail here, could be applied to
decribe the scale symmetry breaking 
effects in gauge field theories.

In the next section we summarize some relevant features of Jordanian
deformation of $sl(2)$ algebraic structure. In section 3 the QYBE with $R$
operators of the fundamental and  generic representations are considered. In
section 4 the realization of representations by polynomials and of
generators by differential operators are formulated. Tensor product
representations are formulated in terms of polynomials in section 5. 
The universal $R$ opertor in spectral and integral forms is considered in
section 6.

\section{Algebraic deformations}
\setcounter{equation}{0}

\subsection{Deformations induced by classical Yang-Baxter solutions}

We start from {\sl Yang}'s solution of QYBE with $sl(2)$ symmetry
\be
\label{Yang}
R^{(0)} (u) = u\ I + \eta P = u  \left ( I +
\eta \ r^{(0)} (u) \right )  \ee
with the corresponding solution of the classical Yang-Baxter equation (CYBE),
\be
\label{r0}
r^{(0)} (u)  = \frac{1}{2u} (I + c_2 ), \ \ \ 
c_2 = \frac{1}{2}
(S_0^+ \otimes S_0^- + S_0^- \otimes S_0^+ + 2
S_0^{0} \otimes S_0^0 ),
\ee
and the Yangian algebraic structure \cite{DrinfeldY,DrinfeldT} associated to them.
$S_0^a$ are the $sl(2)$ generators obeying the standard commutation
relations,
\be
\label{S0}
[S_0^0, S_0^{\pm} ] \ = \ \pm S_0^{\pm}, \ \ \
[S_0^+, S_0^-] \ = \ 2 S_0^0.
\ee

Deformations of this structure are induced by spectral parameter
independent solutions of CYBE. The standard $q$-deformation is induced
by 
\be
\label{rq}
r_q =  S_0^+ \otimes S_0^- - S_0^- \otimes S_0^+
\ee
This deformation can be characterized by the deformed generators
obeying
\be
\label{Sq}
[S_q^0, S_q^{\pm}] = \pm S_q^{\pm}, \ \ \  [S_q^+, S_q^-] = [2 S_q^0]_q,
\ee
with the notation $[a]_q = (q-q^{-1})^{-1} (q^{a} - q^{-a} ) $,
and the deformed co-products, defined on them as
\be
\Delta_q (S_q^{\pm}) = S_{q,1}^{\pm} q^{S_{q,2}^0} +  S_{q,2}^{\pm}
q^{-S_{q,1}^0}, \ \
\Delta_q(S_q^0) = S_{q,1}^0 + S_{q,2}^0, \ \
\bar \Delta_q = \Delta_{q^{-1}}.
\ee
The remarkable feature is here that the co-product for $S^0$ and also the
commutators with it are not changed. From the point  of view of
representations of conformal symmetry in 1 dimension one can say that
the $q$-deformation preserves dilatation symmetry but the
symmetry of translation and proper-conformal transformations is broken.

The classification of solution of CYBE associated with a simple {\sl Lie}
algebra results in a classification of possible deformations \cite{BD}.
In the case of $sl(2)$ there is a further interesting but not so well
studied deformation induced by
\be
\label{rxi}
r_{\xi} = S_0^0 \otimes S_0^{-} - S_0^- \otimes S_0^0 ,
\ee
built from the Borel subalgebra $B_-$ generated by $S_0^-, S_0^0$.
This deformation can be characterized by the deformed generators obeying
\bea
\label{Sxi}
[S_{\xi}^+, S_{\xi}^-] = 2 S_{\xi}^0, \ \ \ \ [S_{\xi}^0, S_{\xi}^-] =
-\frac{1}{\xi} \sh (\xi S_{\xi}^-), \cr
 [S_{\xi}^0, S_{\xi}^+] = \frac{1}{2} \{\ch(\xi S_{\xi}^-), S_{\xi}^+
\},
\eea
and by the co-products defined on them as
\bea
\label{Deltaxi}
\Delta_{\xi} (S_{\xi}^{a}) = S_{\xi,12}^a =
S_{\xi,1}^a  e^{\xi S_{\xi,2}^-} + e^{-\xi S_{\xi,1}^-} S_{\xi,2}^a,
\ \ {\rm for} \ \ a= 0, +, \cr
\Delta_{\xi}(S_{\xi}^- ) = S_{\xi,12}^- = S_{\xi,1}^- + S_{\xi,2}^-,
\ \ \ \ \bar \Delta_{\xi} = \Delta_{(-\xi)}.
\eea
Now the co-products of $S^-$ are not changed. In the conformal symmetry
representations $S_0^-$ can play the role of the translation generator
and one can say that the $\xi$ deformation preserves translation
symmetry whereas the symmetries of dilatation and proper-conformal
transformations are now broken. In this context the deformation
parameter $\xi$ has the natural dimension of length, whereas the
deformation paramenter $q$ is dimensionless.

The $\xi$-deformation of the Casimir operator, i.e. the element of the
elveloping algebra commuting with all $S_{\xi}^a$ and approaching the
$sl(2)$ Casimir operator at $\xi \rightarrow 0$, has the form
\be
\label{Casimir}
{C_{\xi}}=\frac2\xi\sh\frac{\xi S_{\xi}^-}2\;
S_{\xi}^+\;\ch\frac{\xi S_{\xi}^-}2+S_{\xi}^0(S_{\xi}^0+1)
=\frac2\xi\ch\frac{\xi S_{\xi}^-}2\; S_{\xi}^+\;\sh\frac{\xi
S_{\xi}^-}2+S_{\xi}^0(S_{\xi}^0-1) 
\ee

The commutativity with $S_{\xi}^a$ and the different forms are checked
by using the following consequences of (\ref{Sxi}),

\be
\label{Sxif}
S_{\xi}^0e^{\alpha\xi S^-}=e^{\alpha\xi
S_{\xi}^-}\left(S_{\xi}^0-\alpha\sh\xi S_{\xi}^-\right), 
\ee
and
$$
S_{\xi}^+e^{\alpha\xi S_{\xi}^-}=e^{\alpha\xi S_{\xi}^-}\left(S_{\xi}^+
+2\alpha\xi S_{\xi}^0-2\alpha^2\xi\sh\xi S_{\xi}^-\right).
$$

\subsection{Drinfeld twist}

The $\xi$-deformation can be described relying on a general result by
Drinfeld \cite{DrinfeldT}.

Notice that $r_{\xi} \in B_-\otimes B_- $. There exists an element
$F_{\xi}$ in  the elveloping algebra $U(B_-) \otimes
U(B_-)$ such that
\be
\label{FDelta}
F^{12}_{\xi}  ((\Delta_0 \otimes id) F_{\xi} =
F_{\xi}^{23} (id \otimes \Delta_0) F_{\xi}
\ee
($\Delta_0$ is the original co-commutative coproduct) with
\be
\label{Fdecomposed}
F_{\xi} = I \otimes I + \xi r_{\xi} + {\cal O}(\xi^2),
\ee
and such that the deformed co-product can be written as
\be
\label{FDeltaF}
\Delta_{\xi} = F_{\xi} \ \Delta_0 \ F_{\xi}^{-1}.
\ee
If $R^{(0)}$ is a solution of QYBE related to the co-product $\Delta_0$
then
\be
\label{FRF}
R_{12}^{(\xi)} = F_{\xi, 21} \ \ R_{12}^{(0)} \ \ F_{\xi,12}^{-1}
\ee
is a solution of QYBE related to $\Delta_{\xi} $, namely
\be
\label{RxiDelta}
\Delta_{\xi} \ R^{(\xi)} \ = \ R^{(\xi )} \ \bar \Delta_{\xi}.
\ee
The twist element of the $\xi$-deformation of the $sl(2)$ Yangian
can be written in terms of
$$ \sigma_{\xi} = - \ln( I- 2\xi S_0^-) $$
as
\be
\label{Fsigma}
F_{\xi} = \exp(S_0^0 \otimes \sigma_{\xi} ),  \ \ \ 
F_{\xi, 12} = (1 - 2 \xi S_{0,2}^- )^{-S_{0,1}^0}.
\ee
Let us verify (\ref{FDelta}). Rewrite the relation in the sense of
operators acting on $V_1\otimes V_2 \otimes V_3$. Because
$$ (\Delta \otimes id) F = (1 - 2\xi S_{0,3}^-)^{-S_{0,1}-S_{0,2}}, $$
$$
(id \otimes \Delta ) F = (1- 2\xi S_{0,2}^- - 2\xi S_{0,3}^- )^{-S_{0,1}^0}
$$
(\ref{FDelta})  takes the form
\bea
\label{FDelta1}
(1-2\xi S_{0,3}^-)^{S_{0,2}^0}
(1-2\xi S_{0,2}^-)^{-S_{0,1}^0}
(1-2\xi S_{0,3}^-)^{-S_{0,2}^0}
(1-2\xi S_{0,3}^-)^{-S_{0,1}^0}  = \cr
 (1- 2\xi S_{0,2}^- - 2\xi S_{0,3}^- )^{-S_{0,1}^0}
\eea
We notice that
$$
A^{S_{0,2}^0} S_{0,2}^- A^{-S_{0,2}^0}  = A^{-1} S_{0,2}^-, $$
provided $A$ commutes with $S_{0,2}^-$, therefore
$$ (1-2\xi S_{0,3}^-)^{S_{0,2}^0} (1-2\xi S_{0,2}^-)^{-S_{0,1}^0}
(1-2\xi S_{0,3}^-)^{-S_{0,2}^0} =
(1- {2 \xi S_{0,2}^- \over 1-2\xi S_{0,3}^-} )^{-S_{0,1}^0} = $$
$$
 (1- 2\xi S_{0,2}^- - 2\xi S_{0,3}^- )^{-S_{0,1}^0}
(1-2\xi S_{0,3}^-)^{S_{0,1}^0},
$$
which implies (\ref{FDelta1}) and therefore (\ref{FDelta}).

\section{From the fundamental to generic representations}
\setcounter{equation}{0}

\subsection{Fundamental representation R matrix}

We label the representations by the conformal weight $\ell$, the eigenvalue
of $S^0$ of the lowest weight state in the representation space. The
finite-dimensional representations of spin and angular momentum correspond
to negative half-integer and negative integer values of $\ell$. 

Representing the generators by {\sl Pauli} matrices,
\bea
\label{Pauli}
 S_{\xi, -\frac{1}{2}}^+ = \hat  \sigma^+ = \left ( \matrix { 0 & 1 \cr
                                1 & 0  \cr }
              \right ), \ \ \
 S_{\xi, -\frac{1}{2}}^- = \hat \sigma^- = \left ( \matrix { 0 & 0 \cr
                                1 & 0  \cr }
              \right ), \ \ \
S_{\xi, -\frac{1}{2}}^0 = \frac{1}{2}
\hat \sigma^0 = {\frac 12} \left ( \matrix { 1 & 0 \cr
                             0    & -1  \cr }
              \right ),
\eea
the effect of the deformation disappears in the commutation relations
(\ref{Sxi}) and is only present in the co-product (\ref{Deltaxi}). We shall
obtain the solution of QYBE in the fundamental representation
$V_1^{-\frac{1}{2}} \otimes V_2^{-\frac{1}{2}}$.
First we start from the direct implication of (\ref{RxiDelta}),
\be
\label{RS}
R^{(\xi)}_{12} S_{-\xi,12}^a = S_{+\xi,12}^a R^{(\xi)}_{12}.
\ee
The condition with $a=-$ implies that the $4 \times 4 $ matrix
representing R has the form
$$
R^{(\xi)} = u
\left ( \begin{array}{cccc}u+\eta&&&\\r_{21}&u&\eta&\\r_{31}&\eta&u&\\
r_{41}&r_{42}&r_{43}&u+\eta \end{array}
\right ).
$$
The up to now unknown matrix elements vanish at $\xi = 0 $ and in this
way the undeformed solution (\ref{Yang}) is recovered. One of the
remaining conditions, $a=0$ or $a=+$, restricts the ansatz to the
solution,
\be
\label{Rxi}
R^{(\xi),(-\frac{1}{2},-\frac{1}{2})} \ = \ u
\left(\begin{array}{cccc}1&&&\\-\xi&1&&\\\xi&&1&\\\xi^2&-\xi&\xi&1
\end{array} \right)+\eta\left(\begin{array}{cccc}1&&&\\&&1&\\&1&&\\&&&1
\end{array}\right),
\ee
and one checks easily that QYBE,
 \be
\label{fybe} R^{a_1a_2}_{b_1b_2}(u-v)
R^{b_1a_3}_{c_1b_3}(u-w)R^{b_2b_3}_{c_2c_3}(v-w)=
R^{a_2a_3}_{b_2b_3}(v-w)
R^{a_1b_3}_{b_1c_3}(u-w)R^{b_1b_2}_{c_1c_2}(u-v),
\ee
is fulfilled.

In the above arguments we did not use the full algebraic information
available. Using the twist relation (\ref{FRF}) allows to obtain this
result easier. In the representation $V_1^{-\frac{1}{2}} \otimes
V_2^{-\frac{1}{2} }$ the twist element (\ref{Fsigma}) is calculated as
\bea
\label{Ffud}
F_{\xi,12} = (I- 2 \xi S_{0,2}^-)^{-S_{0,1}^0} =   I_{4\times 4} + \xi
\left ( \matrix { 1 & 0 \cr
                                0 & -1  \cr }
              \right ) \
\otimes
\left ( \matrix { 0 &  0 \cr
                                1 & 0  \cr }
              \right ) \
= 
\left ( \matrix { 1 & 0& 0& 0   \cr
                  \xi & 1 & 0 & 0      \cr
                  0 & 0  & 1 & 0  \cr
                  0& 0&  -\xi & 1  \cr}
              \right ). 
\eea
Applying the twist relation (\ref{FRF}),
\be
\label{RFRfund}
R^{(\xi), (-\frac{1}{2} -\frac{1}{2})}_{12} =
F_{\xi,21} \ (u I_{4\times 4}  + \eta P_{4\times4}) \ F_{\xi,12}^{-1} =
u
\ F_{\xi,21} \ F_{\xi,12}^{-1} + \eta P,
\ee
we recover the result (\ref{Rxi}).

\subsection{Lax operator}

We consider QYBE with the representations specified as the
fundamental ones, $\ell_1 = \ell_2 = -\frac{1}{2}$, for the tensor
factors 1 and 2 but generic, $\ell_3 = \ell$, for the 3rd factor
and look for the expression for $R^{(\xi)}_{13}$ or $R^{(\xi)}_{23} $ as a
$2 \times 2$ matrix with the elements being operators in
$V_3^{\ell}$, 
\be 
\label{ansatzLax} 
R^{(\xi),(-\frac{1}{2}, \ell)}(u) 
 = L(u) = \left ( \matrix { a & b \cr
                                c & d  \cr }\right).
\ee 
The $a=-$ component of the symmetry conditions (\ref{RS}) in
the case $\ell_1=-\frac 12$, $\ell_2=\ell$ can be written
explicitly as

 \be\label{b1}
\left(\ba{cc}a&c\\
b&d\ea \right)\left(\ba{cc}S_{\xi}^-&0\\ 1&S_{\xi}^-\ea\right)=
\left(\ba{cc}S_{\xi}^-&0\\
1&S_{\xi}^-\ea\right)\left(\ba{cc}a&c\\ b&d\ea\right), \ee
 and implies
\be
\label{b11}
[b,S_{\xi}^-]=0,\quad [a,S_{\xi}^-]=-b,\quad
[d,S_{\xi}^-]=b,\quad [c,S_{\xi}^-]=a-d.
\ee
We write explicitly also the $a=0$ component of (\ref{RS})
\be
\label{b2}
\left(\ba{cc}a&c\\b&d\ea \right)\left(\ba{cc}S_{\xi}^0+\frac
12e^{\xi S_{\xi}^-}&0 \\
-\xi S_{\xi}^0&S_{\xi}^0-\frac 12e^{\xi S_{\xi}^-}\ea\right)=
\left(\ba{cc}
S_{\xi}^0+\frac 12e^{-\xi S_{\xi}^-}&0\\
\xi S_{\xi}^0&S_{\xi}^0-\frac 12e^{-\xi S_{\xi}^-} \ea\right)
\left(\ba{cc}a&c\\
b&d\ea\right).
\ee
In particular this implies 
\be
\label{b21} 
b(S_{\xi}^0 -\frac
12e^{\xi S_{\xi}^-})= (S_{\xi}^0+\frac 12e^{-\xi S_{\xi}^-})b. 
\ee
The first relation of (\ref{b11}) implies that $b=b(\xi S_{\xi}^-)$ and
(\ref{b21}) then leads to:
$$
b^\prime(\xi S_{\xi}^-)\frac 1\xi\sh\xi S_{\xi}^-=b(\xi S_{\xi}^-)
\frac 1\xi\ch\xi S{\xi}^-
$$
or $b=\frac 1\xi\sh\xi S_{\xi}^-$. We have used the relation
$$F(\xi S_{\xi}^-)S_{\xi}^0=S_{\xi}^0F(\xi S_{\xi}^-)+
F^\prime(\xi S_{\xi}^-)\frac 1\xi\sh\xi
S_{\xi}^-$$ 
which follows from the deformed commutation relations and 
a particular case of  of which is (\ref{Sxif}). With this
result for $b$ the remaining relations (\ref{b11}) imply
 $a=S_{\xi}^0+f(\xi S_{\xi}^-)$,
$d=-S_{\xi}^0+g(\xi S_{\xi}^-)$, $c=S_{\xi}^++x(g-f)+\xi h(\xi
S_{\xi}^-)$, where $f$, $g$ and $h$ are arbitrary functions of
$\xi S_{\xi}^-$ and $x$ obeys the condition $[x, S_{\xi}^-] = -1$.
The remaining conditions involved in (\ref{b2}) are
$$
a(S_{\xi}^0+\frac 12e^{\xi S_{\xi}^-})-\xi bS_{\xi}^0=
(S_{\xi}^0+\frac 12e^{-\xi S_{\xi}^-})a,
$$
$$
d(S_{\xi}^0-\frac 12e^{\xi S_{\xi}^-})=\xi S_{\xi}^0 b+
(S_{\xi}^0-\frac 12e^{-\xi S_{\xi}^-})d,
$$
$$
c(S_{\xi}^0+\frac 12e^{\xi S_{\xi}^-})-\xi dS_{\xi}^0=
(S_{\xi}^0-\frac 12e^{-\xi S_{\xi}^-})c+\xi S_{\xi}^0 a,
$$
and we have four further relations
$$
ae^{\xi S_{\xi}^-}+bS_{\xi}^+=S_{\xi}^+ b+ e^{-\xi S_{\xi}^-}d,
$$
$$
aS_{\xi}^+-\xi bS_{\xi}^+=S_{\xi}^+a+e^{-\xi S_{\xi}^-}c,
$$
$$
ce^{\xi S_{\xi}^-}+dS^+=\xi S_{\xi}^+ b+S_{\xi}^+d,
$$
$$
cS_{\xi}^+-\xi dS_{\xi}^+=S_{\xi}^+c+\xi S_{\xi}^0 c,
$$
which follow from the $a=+$ component of the symmetry relations
({\ref{RS}).
They are sufficient to obtain the explicit form of functions $f$, $g$
and $h$ and this leads to the result for the Lax operator, 
\be
\label{Lxi} 
L^{(\xi)} (u)=L^{(\xi)} + u l^{(\xi)}, \ee
$$
L^{(\xi)} \equiv\left(
\begin{array}{cc}
\frac 12\ch\xi S_{\xi}^-+S_{\xi}^0&\frac 1\xi\sh\xi S_{\xi}^-\\
S_{\xi}^+&\frac 12\ch\xi S_{\xi}^--S_{\xi}^0
\end{array}
\right),\quad
l^{(\xi)} \equiv\left(
\begin{array}{cc}
e^{-\xi S_{\xi}^-}&0\\
2\xi S_{\xi}^0+\xi\sh\xi S_{\xi}^-&e^{\xi S_{\xi}^-}
\end{array}
\right).
$$
The resulting Lax operator obeys  QYBE in the form
\be
\label{ybe}
R_{12}(u-v)L_1(u)L_2(v)=L_2(v)L_1(u)R_{12}(u-v) \ee
due to the intertwining relation with the co-product symbols
(\ref{RxiDelta}, \ref{RS}), and this has been checked by direct but tedious
calculations in the following way.
 We represent  the fundamental $R$-matrix (\ref{Rxi})
and Lax operator (\ref{Lxi}) as follows:
$$
R_{ab}^{cd}(u)=(\delta_1)_a^c(\delta_2)_b^d-
\xi(\sigma^0_1)_a^b(\sigma_2^-)_b^d+
\xi(\sigma^-_1)_a^b(\sigma^0_2)_b^d+
\xi^2(\sigma_1^-)_a^b(\sigma^-_2)_b^d,
$$
$$
\!\!\!\!\!(L_i(u))_a^b\!=\!(u+\frac 12)\ch\xi
S_{\xi}^-(\delta_i)_a^b+[S_{\xi}^0\!-u\sh\xi
S_{\xi}^-](\sigma^0_i)_a^b+\frac 1\xi\sh\xi S_{\xi}^-(\sigma^+)_a^b+
[S_{\xi}^+\!+u\xi(2S_{\xi}^0+\sh\xi S_{\xi}^-)](\sigma^-_i)_a^b,
$$
 and multiplying Pauli matrices acting in the same space, we
obtain on both sides of QYBE an expansion
over terms of the type $\sigma_1^a\otimes\sigma_2^b, a,b = \pm,0$
acting in  $V_1^{-\frac 12}\otimes V_2^{-\frac 12}$
multiplied with operators acting in
$ V_3^\ell$. Then we check that the operator-valued coefficients
of the corresponding  $\sigma_1^a\otimes\sigma_2^b$ term on both sides are
equal by using the commutation relations (\ref{Sxi}) of the deformed
generators.

We have the alternative option for constructing the Lax operator by the
twist relation (\ref{FRF}). It reads in the case at hand
\be
\label{FLF}
L^{(\xi)} (u) = F_{\xi,21} \ L^{(0)} (u) \ F_{\xi,12}^{-1},
\ee
with the well known undeformed Lax operator (with $u$ shifted by $\frac
12$)
\bea
\label{L0}
L^{(0)}(u)\! = \!\left(\!
\begin{array}{cc}
u+\frac 12+S_0^0&S_0^-\\
S_0^+&u+\frac 12-S_0^0
\end{array}\!
\right)\!.
\eea
We write the twist element in the representation $\ell_1 = -
\frac{1}{2}, \ell_2 = \ell$,
\bea
F_{\xi,12} = \exp(\frac{1}{2} \hat \sigma^0_1 \otimes \sigma_{\xi,2} ),
\ \ e^{-\sigma_{\xi,2}} = 1 - 2 \xi S_{0,2}^-, \cr
F_{\xi,21} = \exp(S_{0,2}^0 \otimes \hat \sigma_{\xi,1} ), \ \ \ \
e^{-\sigma_{\xi,1}} = \ I - 2 \xi \hat \sigma^-.
\eea
Here Pauli matrices $\hat \sigma^a$ are to be distinguished from the
two representations of the operator $\sigma_{\xi}$. The representation
label $\ell$ on the operators acting in $V^{\ell} $ is suppressed and will
be written as an additional subscript if necessary.
Now the twist elements can also be written as $2\times 2$ matrices with
elements acting as operators in $V^{\ell}$,
$$
{F}_{12}=\left(\ba{cc}
e^{\sigma_{\xi}/2}&\\&e^{-\sigma_{\xi}/2}\ea\right)=
\left(\ba{cc}(1-2\xi S_0^-
)^{-\frac 12} &0\\0&(1-2\xi S_0^-)^{\frac 12}\ea\right),
$$
\be
\label{telement}
{F}_{21}=\left(\ba{cc} 1&0\\ 2 \xi S_0^0
   &1\ea\right).
\ee

As the result we obtain the Lax operator in terms of the undeformed
generators $S_0^a$ in the generic representation $\ell$,
$$  L^{(\xi)}(u)\! = $$
\bea
\label{Lxi0}
 \!\left(\!
\matrix {
(u+\frac 12+S_0^0)(1-2\xi S_0^-)^{\frac 12} & S_0^- (1- 2\xi S_0)^{-\frac
12} \cr
[(u+ \frac 12 +S_0^0) 2 \xi S_0^0+S^+] (1-2 \xi S_0^-)^{\frac
12} &  [ 2 \xi S_0^0 S_0^- +u+\frac 12-S_0^0] (1 - 2\xi S_0^-)^{-\frac 12}
}  \right)\!.
\eea
The comparison of both results (\ref{Lxi}, \ref{Lxi0}) leads to the
following expressions of the deformed generators $S_{\xi}^a $ in terms of
the undeformed ones $S_0^a$,
\bea
\label{SxiS0}
S_{\xi}^- = - \frac{1}{2\xi} \ln(1- 2\xi S_0^-), \cr
S_{\xi}^0 = S_0^0 (1-2 \xi S_0^-)^{\frac 12} - \frac{\xi}{2} S_0^-
(1-2\xi S_0^-)^{-\frac 12}, \cr
S_{\xi}^+ = [S_0^+ + 2 \xi S_0^0 (S_0^0 + \frac 12)] (1-2\xi
S_0^-)^{\frac 12}.
\eea
With these expressions one can prove by direct computation the
commutation relations (\ref{Sxi}) assuming that $S_0^a$ obey the
undeformed $sl(2)$ algebra relations (\ref{S0}).
This calculation is outlined in Appendix A.

There is another relation of the deformed generators $S_{\xi}^a$
to the undeformed
$sl(2)$ algebra \cite{Jeugt}. The three functions of $S_{\xi}^a$,
\be
\label{Jeugt}
S_J^- = {\sh(\xi S_{\xi}^-) \over \xi (\ch(\frac{\xi}{2} S_{\xi}^-) )^2 },
\ \ S_J^0 = S_{\xi}^0, \ \ 
S_J^+ = \ch(\frac{\xi}{2} S_{\xi}^-) \ S_{\xi}^+ \ \ch(\frac{\xi}{2}
S_{\xi}^-),
\ee
obey the undeformed commutation relations (\ref{S0})
(with $S_0^a$ replaced by $S_J^a$).

\subsection{Universal R operator}

We consider QYBE for generic $\ell_1$ and $\ell_2$ and $\ell_3 =
-\frac{1}{2}$. It involves the universal R operator, $R^{(\xi,
\ell_1,\ell_2)} (u)_{12}$ and $R_{i3}^{(\xi, \ell_i, -\frac{1}{2})} , \ i=1,2$
which can be represented by the Lax matrices just obtained (\ref{Lxi}),
\be
\label{RL1L2}
R^{(\xi)}_{12} (u) \ L^{(\xi)}_1 (u+v) \ L_2^{(\xi)} (v) =
L_2^{(\xi)}(v)  L_1^{(\xi)} (u+v) \ R^{(\xi)}_{12} (u).
\ee
The Lax operators involve the generators acting on space 1 or 2
correspondingly in the representations $\ell_1 $ or $\ell_2$. This $2 \times
2$ matrix  relation
serves as the defining condition for the universal R operator acting on
$V_1^{\ell_1} \otimes V_2^{\ell_2}$.
By decomposition in powers of $v$ we obtain three conditions. The first two,
\be
\label{sym}
R_{12}(u)l_1l_2=l_2l_1R_{12}(u),
\ee
$$
R_{12}(u)(L_1l_2+l_1L_2)=(L_2l_1+l_2L_1)R_{12}(u),
$$
are equivalent to the symmetry relations (\ref{RS}), now for generic
representations. The third condition reads
\bea
\label{ybeu}
R_{12}(u) K_L(u)  = K_R (u) R_{12} (u), \cr
K_L (u) =
L_1L_2+\frac u2(l_1L_2-L_1l_2)  , \cr
K_R(u) =
L_2L_1+\frac u2(L_2l_1-l_2L_1).
\eea
 The elements of the $2 \times 2$ matrices $K_{L/R}$ are
operators on $V_1^{\ell_1}\otimes V_2^{\ell_2}$
and  their explicite form is listed in Appendix B.,

Together with the symmetry conditions (\ref{RS}) one of the four  conditions
involved in the $2 \times 2$ matrix relation (\ref{ybeu}) is sufficient to fix
the operator $R^{(\xi)}_{12} (u)$ up to normalization.

From the deformed commutation relations (\ref{Sxi})) we see that $S_{\xi}^a$
may be represented in a form even in the deformation variable $\xi$. The
dependence on $\xi$ is not on even powers only for the form (\ref{SxiS0})
with the $S_0^a$ independent on $\xi$. For $S_{\xi}^a$ symmetric in $\xi$
the second term in $ L^{\xi} (u) = L^{(\xi)} + u l^{(\xi)}$ (\ref{Lxi})
is not symmetric with respect to $\xi
\leftrightarrow - \xi $. Because
\be
( L^{(\xi)} )^2 = (C_{\xi} + {\frac 14}) I,
\ \ l^{(-\xi)} = (l^{(\xi)})^{-1}, \ \
l^{(\xi)} L^{(\xi)} = L^{(\xi)} l^{(-\xi)},
\ee
we find that
\be
L^{(-\xi)} (-u) = (C_{\xi} + \frac{1}{4} - u^2) \ (L^{(\xi)} (u))^{-1},
\ee
i.e.
on a given irreducible representation the inverse of $L^{(\xi)} (u)$
is proportional to
$L^{(-\xi)} (-u)$. This property holds also for the universal R operator
$ R^{(\xi, \ell_1,\ell_2)} $ in accordance with the two forms of QYBE,
the above one (\ref{RL1L2}) and the following one,
\be
\label{RL2L1}
R^{(-\xi)}_{21} (u) \ L_2^{(-\xi)} (u+v)_1 \ L_1^{(-\xi)} (v) =
L_2^{(-\xi)}(v)  L_2^{(-\xi)} (u+v) \ R^{(-\xi)}_{12} (u).
\ee

We notice that the Lax operator can be made symmetric in $\xi$ (for
symmetric $S_{\xi}^a$ ) by the following similarity transformation,
\bea
\label{Lxisym}
L^{(\xi) sym} (u) = e^{u \xi S_{\xi,1}^-} \ L^{(\xi)}(u) \  e^{-u \xi
S_{\xi, 1}^-} = \cr
\left(
\begin{array}{cc}
\frac 12\ch\xi S_{\xi}^-+S_{\xi}^0&\frac 1\xi\sh\xi S_{\xi}^-\\
S_{\xi}^+&\frac 12\ch\xi S_{\xi}^--S_{\xi}^0
\end{array}
\right)+
u\left(
\begin{array}{cc}
\ch{\xi S_{\xi}^-}&0\\
(u+1)\xi\sh\xi S_{\xi}^-&\ch{\xi S_{\xi}^-}
\end{array}
\right).
\eea
The resulting matrix operator obeys QYBE with the fundamental representation
R matrix (\ref{fybe}) as well as the original one. Taking into account that the
universal R operator commutes with $S_{\xi,1}^- + S_{\xi,2}^- $ we obtain
that
\be
\label{Rxisym}
R^{(\xi) sym}_{12} (u) = e^{u \xi S_{\xi}^-} \ \ R_{12}^{(\xi)} (u) \ \
 e^{-u \xi S_{\xi}^-},
\ee
obeys (\ref{RL1L2}) with $L^{(\xi) }$ replaced by $L^{(\xi)sym }$
and is therefore symmetric in $\xi \leftrightarrow - \xi $.

\section{Representations by polynomials of one variable}
\setcounter{equation}{0}

Guided by the motivations described in the Introduction we would like to
represent the considered algebras in terms of operators acting on polynomial
functions. We consider the representations with one variable in this section
and the tensor product representations involving two varibles in the next
section.

\subsection{Representations of the deformed algebra}

Recall that the undeformed generators (\ref{S0}) can be represented 
on functions $\phi(y)$ as differential operators,
\be
\label{S0y}
S_{0, \ell}^- = \dd_y, \ \ S_{0,\ell}^0 = \frac{1}{2} (y \dd_y + \dd_y y) +
\ell - \frac{1}{2}, \ \
S_{0,\ell}^+ = - y \dd_y y - (2 \ell -1
) y.
\ee
Representations with conformal weight $\ell$ are spanned by the polynomials
\be
\phi_{\ell}^{(m)} (y) = (S_{0,\ell}^+)^m \phi_{0,\ell}^{(0)} (y),
\ee with the lowest weight state represented by
$ \phi_{0,\ell}^{(0)} (y) = 1$. Up to normalization this basis is given by
the monomials, $\phi_{\ell}^{(m)} (y) =  y^m $.
In the particular case $\ell = 0$ $S_{0,0}^a$ generate the {\sl M\"obius}
transformations of the variable $y$,
$$ y \rightarrow y\p = { a y + b \over c y + d}.$$

The deformed generators (\ref{Sxi}) can be represented as
\be
\label{Sxix}
S_{\xi, \ell}^- = \dd_x, \ \ S_{0,\ell}^0 = \frac{1}{2} (x \dd_x^{\xi}
+ \dd_x^{\xi} x) + \ell - \frac{1}{2}, \ \
S_{\xi,\ell}^+ = - x \dd_x^{\xi} x - (2 \ell -1) x,
\ee
where
$$ \dd_x^{\xi} = \frac{1}{\xi}  \sh(\xi \dd_x). $$
Substituting these generators into the expression of the Casimir operator
(\ref{Casimir}) we
confirm that the restriction of $C_{\xi}$ to the representation with weight
$\ell$ is the identity operator times $\ell (\ell-1)$.

If we would substitute the representations (\ref{S0y}) for $S_{0,\ell}^a$ to
(\ref{SxiS0}) we would obtain a quite different representation of the
deformed generators in terms of $(y,\dd_y)$. That representation is more
complicated. First of all it has the disadvantage that $S_{\xi,\ell}^-$ does
not appear as the infinitesimal translation in $y$, further, it is not
symmetric in $\xi$. .

The lowest weight states of the deformed representation $\ell$
obey
\be
\label{S-phi0}
S_{\xi,\ell}^- \ \varphi_{\ell}^{(0)} (x) = 0, \ \ \
S_{\xi,\ell}^0 \ \varphi_{\ell}^{(0)} (x) = \ell  \varphi_{\ell}^{(0)} (x),
\ee
and they are again represented by the constant functions,
$\varphi_{\ell}^{(0)} (x) = 1 $. The basis polynomials of this
representation obeying the eigenvalue condition
\be
\label{S0phim}
S_{\xi,\ell}^0 \ \varphi_{\ell}^{(m)} = (\ell + m) \ \varphi_{\ell}^{(m)}
\ee
cannot be obtained by applying $S_{\xi,\ell}^+$ to the constant. In order to
construct these basis polynomials we consider the {\sl van der Jeugt}
operators $S_J^a$ (\ref{Jeugt}).
Using the representation (\ref{Sxix} ) we can write them
in terms of the following {\sl Heisenberg} canonical pair,
\be
\label{Jeugtpair}
X_J = \ch(\frac{\xi}{2} \dd_x) x \ch(\frac{\xi}{2} \dd_x), \ \
D_J = (\ch(\frac{\xi}{2} \dd_x)^{-2} \dd_x^{\xi}, \ \
[D_J, X_J] = 1,
\ee
in the form
\be
\label{SJx}
S_{J, \ell}^- = D_J, \ \ S_{J,\ell}^0 = \frac{1}{2} (X_J D_J + D_J X_J) +
\ell - \frac{1}{2}, \ \
S_{J,\ell}^+ = - X_J D_J X_J - (2 \ell -1) X_J.
\ee
$\phi_{\ell}^{(m)} (X_J)$ span the representation $\ell$ of the undeformed
algebra generated by $S_{J,\ell}^a$. On the other hand, considering the
generators $S_{J,\ell}^a$ as operators on polynomial functions of $x$,
the conditions
(\ref{S-phi0}, \ref{S0phim}) are equivalent to
\be
\label{SJphi}
S_{J,\ell}^- \ \varphi_{\ell}^{(0)} (x) = 0, \ \ \
S_{J,\ell}^0 \ \varphi_{\ell}^{(m)} (x) = (\ell + m) \
\varphi_{\ell}^{(m)} (x),
\ee
We obtain for the basis polynomials of the deformed representation
\be
\label{philmx}
\varphi_{\ell}^{(m)} (x) = (S_{J,\ell}^+)^m \cdot 1 = const \ X_J^m \cdot 1.
\ee
The explicite polynomials are obtained by substituting $X_J$ in terms of
$(x,\dd_x)$ (\ref{Jeugtpair}) and commuting the differential operators to
the right. Specifying the normalization to be such that the coefficient of
the highest power is 1 we have e.g.
$$
\varphi^{(0)}=1,\quad
\varphi^{(1)}=x,\quad
\varphi^{(2)}=x^2+\frac 14,\quad
\varphi^{(3)}=x^{3}+\frac 54x,\quad
\varphi^{(4)}=x^{4}+\frac 72x^2+\frac{9}{16},
$$
\be
\label{pols}
\varphi^{(5)}=x^5+\frac {15}{2}x^3+\frac{89}{16}x,\quad
\varphi^{(6)}=x^6+\frac{55}{4}x^4+\frac{439}{16}x^2+\frac{225}{64}.
\ee

The deformed basis polynomials $\varphi^{(m)}_{\ell}$ 
can be computed by the following rule:
Expand
$$ [(2m_1+1)!!]^2 \ (\frac{\xi^2}{4} + p)^{m_1} $$
in powers of $p$ and substitute
for even $m= 2m_1+2$,
$$p^k \rightarrow x \prod_{r=-k+1}^{k-1} (x + r \xi)
[(2k-1)!!]^{-2},
$$
and for odd $m= 2m_1 +1$
\be
\label{philmrule}
p^k \rightarrow  \prod_{r=-k}^{k} (x + r \xi)
[(2k+1)!!]^{-2}
\ee
The derivation is given in Appendix C.

There is also a generating function for these polynomials
\be
\label{repf}
G_1(x,t)\equiv\sum_{m=0}^\infty\frac{t^m}{m!}\varphi^{(m)}(x)=
\frac{\left(1+\frac{t\xi}2\right)^{\frac x\xi-\frac 12}}
{\left(1-\frac{t\xi}2\right)^{\frac x\xi+\frac 12}}.
\ee
For a derivation we refer to Appendix C.
It can be checked that the generating function
$G_1 (x,t) $ satisfies the following
difference-differential equation,
$$
\frac x{2\xi}\left(G_1(x+\xi,t)-G_1(x-\xi,t)\right)+\frac
14\left(G_1(x+\xi,t)+G_1(x-\xi,t)-2G_1(x,t)\right)=t\frac\dd{\dd t}G_1(x,t),
$$
which is the implication of the eigenvalue condition in (\ref{SJphi})
read as a difference equation for $\varphi_{\ell}^{(m)} $.

\subsection{Induced representation }

The preferred representation of the deformed generators (\ref{Sxix}),
where $S_{\xi}^- = \dd_x$ is the infinitesimal translation, is connected via
the twist (\ref{SxiS0}) to a representation of the undeformed generators
$S_0^a$.

We use a two-parameter family of {\sl Heisenberg} canonical pairs,
\be
\label{xialphapair}
D_x^{(\xi)} = e^{-\xi \dd_x} \ \dd_x^{\xi}, \ \ \
X^{\xi, \alpha} = (1+ e^{-\xi \dd_x} )^{-\alpha} \
e^{\xi \dd_x}  x e^{\xi \dd_x}   (1+  e^{-\xi \dd_x} )^{\alpha},
[D_x^{(\xi)}, X^{\xi, \alpha}] = 1,
\ee
and show that the representation of $S_{0,\ell}^a$ induced by the one of
$S_{\xi,\ell}^a $ (with the $a=-$ component generating translations in $x$)
is given by (\ref{S0y}) with the pair $(y, \dd_y)$ replaced by the pair
$(X^{(\xi, 2 \ell -1)}, D_x^{(\xi)} )$.

Indeed, inverting the first relation of (\ref{SxiS0})
we have
\be
\label{S0-}
S_{0,\ell}^- = \frac{1}{2\xi} \left ( 1- e^{-2\xi S_{\xi, \ell}^-} \right )
= D_x^{(\xi)}, \ \ \ \
\sqrt {1-2\xi S_{0,\ell}^-} = e^{-\xi\dd_x}
\ee
The second relation in  (\ref{SxiS0}) can be written as
$$ S_{0,\ell}^0 = \left ( S_{\xi,\ell}^0 + \frac{\xi}{2} D_x^{(\xi)}
e^{\xi\dd_x} \right ) e^{\xi \dd_x}. $$
Substituting the representation (\ref{Sxix}) we obtain as an intermediate
form
$$S_{0,\ell}^0 = \frac{1}{2} \left ( X^{(\xi,0)} D_x^{(\xi)} +
 D_x^{(\xi)} X^{(\xi,0)} \right )
+ (\ell - {\frac 12}) e^{\xi \dd_x}. $$
Now we substitute
\be
\label{X0alpha}
X^{(\xi,0)} = X^{(\xi,\alpha)} - { \alpha \xi e^{2\xi \dd_x} \over
1+ e^{\xi \dd_x} },
\ee
and obtain for $ \alpha = 2 \ell -1 $
\be
\label{S00}
S_{0,\ell}^0 =
 \frac{1}{2} \left ( X^{(\xi,2\ell -1)} D_x^{(\xi)} +
 D_x^{(\xi)} X^{(\xi,2\ell -1)} \right )
+ \ell - {\frac 12}.
\ee
The third relation in (\ref{SxiS0}) leads to
$$ S_{0,\ell}^+ = S_{\xi,\ell}^+ e^{\xi \dd_x} \
 - \ 2\xi
[\frac{1}{2} \left ( X^{(\xi,2\ell -1)} D_x^{(\xi)} +
 D_x^{(\xi)} X^{(\xi,2\ell -1)} \right )
+ (\ell - {\frac 12}) ] \cdot $$
$$
[\frac{1}{2} \left ( X^{(\xi,2\ell -1)} D_x^{(\xi)} +
 D_x^{(\xi)} X^{(\xi,2\ell -1)} \right )
+ \ell  ]. $$
We substitute (\ref{Sxix}) and show that the remainder in
\be
\label{S0+}
S_{0,\ell}^+ =
- X^{(\xi,2\ell -1)} D_x^{(\xi)}   X^{(\xi,2\ell -1)}
- (2\ell -1) X^{(\xi,2\ell -1)} +
(S_{0,\ell}^+)_r
\ee
vanishes, $(S_{0,\ell}^+)_r = 0$.

The undeformed generators are represented now by rather involved difference
operators in $(x, \dd_x)$ and they depend now on the deformation parameter
$\xi$. The basis polynomials of the representation
$\ell$ of the undeformed algebra, being simple monomials in $y$, become
now more involved polynomials in $x$,
\be
\label{phix0}
\phi_{x,\ell}^{(m)} (x) = \phi_{x,\ell}^{(m)} (X^{(\xi, 2 \ell -1)})
\cdot 1 \ = \ \ 2^{-(2\ell-1) m} \
\prod_{k=1}^m (x + (\ell -\frac{1}{2} + 2k-1)\xi ).
\ee
The original form of the undeformed representations (\ref{S0y})
will be referred to as the $y$ picture whereas the induced form
(\ref{S0-}, \ref{S00}, \ref{S0+}) will be called $x$ picture.

\section{Polynomial representations of tensor products}
\setcounter{equation}{0}

\subsection{Undeformed tensor product}

For comparison we recall the undeformed case,   where
$\Delta_0 (S_0^a) = S_{0, 1}^a + S_{0, 2}^a = S_{0,12}^a$ and where
the lowest
weight states of the irreducible representations appearing in the tensor
product of representations $\ell_1$ and $\ell_2$ are represented by the
solutions of
\be
\label{lowestw0}
S_{0,12}^- \phi_{12,\ell_1,\ell_2,n}^{(0)} = 0, \ \ \
S_{012}^0 \phi_{12,\ell_1,\ell_2,n}^{(0)} = (\ell_1 + \ell_2+n) \
\phi_{12,\ell_1,\ell_2,n}^{(0)}.
\ee
We shall usually suppress the labels $\ell_1,\ell_2$.

In the $(y,\dd_y)$ picture (\ref{S0y}) we see, that $\phi_{12,n}^{(0)}$
depends on the difference $y_1 - y_2 = y_{12} $ only and that
\be
\label{phi12ny}
\phi_{12,n}^{(0)} (y_1,y_2)  \ = \ y_{12}^n.
\ee
The states of an irreducible representation with weight $\ell = \ell_1 +
\ell_2 +n$ are generated as
\be
\label{phi12nmy}
\phi_{12,n}^{(m)} (y_1,y_2)  \ = \ (S_{0,12}^+)^m \ \
\phi_{12,n}^{(0)} (y_1,y_2).
\ee

The same undeformed representation can also be described in the $(x,\dd_x)$
picture (\ref{S0-}, \ref{S00}, \ref{S0+}).
The polynomial $\phi_{x,12,n}^{(m)} (x_1,x_2) $ is obtained
from the above one by substituting in the decomposition of (\ref{phi12nmy})
in the monomials $y_1^{m_1} y_2^{m_2} $, according to (\ref{phix0}),
\be
\label{phixnmrule}
 y_i^{m_i} \rightarrow 2^{-(2\ell_i -1)m_i} \ \prod_{k=1}^{m_i}
(x_i + (\ell_i -\frac{1}{2} + 2k-1)\xi).
\ee
In particular the lowest weight polynomials do not depend on  the
difference $x_{12}$ only.

\subsection{Deformed tensor product}

The generators $S_{\xi,12}^a$ acting on $V_1^{\ell_1} \otimes V_2^{\ell_2}
$are given by (\ref{Deltaxi}, \ref{Sxix}) and the conditions on the
polynomials representing the lowest weight states of the irreducible
representations appearing in the deformed tensor product are
\be
\label{lowestwxi}
S_{\xi,12}^- \varphi_{12,\ell_1,\ell_2,n}^{(0)} = 0, \ \ \
S_{\xi,12}^0 \ \varphi_{12,\ell_1,\ell_2,n}^{(0)} = (\ell_1 + \ell_2+n) \
\varphi_{12,\ell_1,\ell_2,n}^{(0)}.
\ee
The labels $\ell_1,\ell_2$ will be suppressed.
Since $S_{\xi,12}^- = \dd_{x_1} + \dd_{x_2}$ we see that here again the
dependence is on the difference $x_1 - x_2 = x_{12} $ only. Using this fact
the second condition can be written in the form
$$ [\dd_{x_1}^{\xi} x_{12} + \ell_1 + \ell_2 -1] e^{-\xi \dd_{x_1}}
 \ \varphi_{12,n}^{(0)} (x_{12}) = (\ell_1 + \ell_2+n) \
\varphi_{12,n}^{(0)}(x_{12}).
$$
We shall solve this eigenvalue equation by expressing the involved operator
in terms of the canonical {\sl Heisenberg } pair (\ref{xialphapair}).
We abrreviate $x_{12}$ by $x$ and then the latter equation can be written as
$$ [D_x^{(-\xi)} X^{(-\xi,0} + (e^{-\xi \dd_x} -1) (\ell_1 + \ell_2 -1) ]
 \ \varphi_{12,n}^{(0)} = (n+1)  \ \varphi_{12,n}^{(0)}.
$$
We notice that
$$e^{\xi \dd_x} -1 = -2 \xi D_x^{(-\xi)} \ { e^{-2\xi \dd_x} \over
1 + e^{\xi \dd_x} }$$
and use  also (\ref{X0alpha}) with $\xi$ replaced by $-\xi$ and $\alpha$ by
$2(\ell_1 + \ell_2 -1)$ to obtain the eigenvalue equation in the form
\be
\label{lowestwxiX}
D_x^{(-\xi)} \ X^{(-\xi, 2(\ell_1 + \ell_2 -1))} \
\varphi_{12,n}^{(0)} = (n+1) \varphi_{12,n}^{(0)}
\ee
In the $(X^{(-\xi, 2(\ell_1 + \ell_2 -1))},D^{(-\xi)})$  picture the
eigenfunctions are just
$$ \tilde  \varphi_{12,n}^{(0)}  = \ \
(X^{(-\xi, 2(\ell_1 + \ell_2 -1))} )^n $$
with $n$ all non-negative integers.
Returning to the $(x,\dd_x)$ picture we obtain the polynomials representing
the lowest weight states in the $\xi$ deformed
tensor product of representations with
conformal weights $\ell_1, \ell_2$ as
\bea
\label{phin0x12}
\varphi_{12,n}^{(0)} = (X^{(-\xi, 2(\ell_1 + \ell_2 -1))} )^n \cdot
1|_{x=x_{12}}
 = \cr
 2^{-(2(\ell_1+\ell_2-1) n} \ \
\prod_{k=1}^n \ (x_{12} - \xi ( \ell_1 + \ell_2 -1 + 2k-1)).
\eea

Actually we have just solved a non-trivial difference equation. The
eigenvalue condition (\ref{lowestwxi}) with $S_{\xi, 12}^0$ acting
on functions of $x = x_{12}$  reads
\be
\label{s0phi}
\frac{x}{2\xi} [\varphi_n( x)-\varphi_n
( x-2 \xi)]+(\ell_1 + \ell_2) \varphi_n( x-\xi)+\frac 12[\varphi_n( x)+
\varphi_n(x-2 \xi)-2\varphi_n( x-\xi)]=
\ee
$$
=(\ell_1 + \ell_2 + n)\varphi_n( x),
$$
We shall denote the left-hand side by $\tilde S_{\xi, x}^0 \varphi_{n}$.

By representing the function by a {\sl Fourier} integral one can deduce
similar to the one-point case (\ref{varpin}) an expression of these
polynomials
$$
\varphi_n( x)=c_n \int dke^{ik( \frac{x}{\xi} +1-\ell_1-\ell_2-n)}
(\sin k)^{-\ell_1-\ell_2-n}
\left(\frac{\cos\frac k2}{\sin\frac k2}\right)^{1-\ell_1-\ell_2}.
$$
and derive from it recursive relations like
$$
(1-n- \ell_1 - \ell_2) \varphi_n( x)-(1 - \ell_1 - \ell_2) \varphi_n(
x-\xi) =   \frac{c_n}{c_{n-1}} i (1 - \frac{x}{\xi})\varphi_{n-1}( x-\xi),
$$
and
\be
\label{phinrec}
\varphi_n( x)-\varphi_n( x-2\xi )=2i
\frac{c_n}{c_{n-1}}  \varphi_{n-1}( x-2 \xi).
\ee
 The explicite expression obtained obove (\ref{phin0x12})
 obeys indeed these relations with the coefficients related as
$$ c_n = -i n  c_{n-1}. $$

The difference equation (\ref{s0phi}) can be solved directly, starting from
the special case $\ell_1 + \ell_2 =1$ where it simplifies to
$$
(\frac{x}{\xi} -1) \frac{1}{2} [\varphi_n( x)-\varphi_n( x-2\xi)]
= n  \varphi_n( x).
$$
and leads to the solution (\ref{phin0x12}) for this particular case.
We summarize this solution for all $n$ into
the generating function
$$
G_2^{(+1)}(x,\tau)= \sum_{n=0}^{\infty} \tau^n \varphi_n (x) =
(1+2\tau \xi)^{\frac{ x-\xi}{2\xi}}.
$$
The equation (\ref{lowestwxi}) in the general case is obtained
from this special one by
similarity transformation with $ (1+ e^{\xi \dd_x})^{-2(\ell_1+\ell_2-1)}$.
This allows to recover the above result (\ref{phin0x12})
and also to represent the
general case lowest weight polynomials by the following generating function
\be
\label{rfell}
G_2^{(\ell_1 + \ell_2)}( x,\tau)=
({1+e^{ \xi\dd_x}})^{-2(\ell_1 + \ell_2 - 1)}
(1+2\tau \xi)^{\frac{ x-\xi}{2\xi} }=(1+2\tau \xi)^{\frac{ x-\xi}{2\xi}}
({1+\sqrt{1+2\tau \xi }} )^{-2(\ell_1 +\ell_2 - 1)}.
\ee

\section{The universal R operator}
\setcounter{equation}{0}

\subsection{The spectral form}

According to (\ref{RxiDelta}) the universal R operator
$R_{12,\ell_1,\ell_2}^{(\xi)}(u)$ maps the irreducible representation
$\ell =\ell_1 + \ell_2 + n$ in the
tensor product $V_1^{\ell_1} \otimes V_2^{\ell_2} $ constructed according
to $\Delta_{\xi} $ into the corresponding one of $\bar \Delta_{\xi} =
\Delta_{-\xi}$ and vice versa. In particular the symmetry relations
(\ref{RS})  guarrantee that the basis functions are mapped into each other.
This is expressed by the following generalized eigenvalue relations,
\bea
\label{Reigenv}
R_{12,\ell_1,\ell_2}^{(\xi)}(u) \ \ \varphi_{12, n}^{(\xi),(m)} (x_1,x_2) =
\rho_{\ell_1 , \ell_2, n} \ \ \ \varphi_{12, n}^{(-\xi),(m)} (x_1,x_2), \cr
R_{12,\ell_1,\ell_2}^{(\xi)}(u) \ \ \varphi_{12, n}^{(-\xi),(m)} (x_1,x_2) =
\bar \rho_{\ell_1 , \ell_2, n} \ \ \ \varphi_{12, n}^{(\xi),(m)} (x_1,x_2).
\eea
By symmetry the eigenvalues $\rho_{\ell_1,\ell_2, n}$ do not depend on the
level $m$. We have pointed out above that $R_{12}^{(\xi)}$ can be made
symmetric in $\xi \leftrightarrow -\xi$ by a similarity transformation. This
implies that the eigenvalues do not depend on the sign of $\xi$,
therefore
$$ \rho_{\ell_1,\ell_2, n} = \bar \rho_{\ell_1,\ell_2, n}. $$
We check now that the above eigenvalue relations are compatible with the
defining conditions of the universal R operator if applied to the lowest
weight basis polynomials $(m=0)$.  It is sufficient to consider the
conditions (\ref{ybeu}) resticted to the upper right $(12)$ element in the
$2\times 2$ matrices,
\be
\label{RKphi}
R_{12,\ell_1,\ell_2}^{(\xi)}(u) K_L^{12} (u) \ \ \varphi_{12, n}^{(-\xi),(m)}
(x_1,x_2)
 \ \ = \ \ K_R^{12} (u)\ \   R_{12,\ell_1,\ell_2}^{(\xi)}(u)
\varphi_{12, n}^{(-\xi),(m)} (x_1,x_2)
\ee
We consider the action of $K_{L/R}^{12}(u)$ as difference operators on
functions of $x_{12} = x$,
\bea
\label{KLRdiff}
2 \xi K_{L/R} \varphi (x) =
\pm \frac{x}{2\xi} [ 2\varphi (x) - \varphi (x + 2\xi) - \varphi (x - 2\xi) ]
\cr
\pm (1-\ell_1-\ell_2) [\varphi (x+\xi) - \varphi (x-\xi) ]  \cr
\mp \frac{1}{2} [\varphi (x + 2\xi) - \varphi ( x - 2\xi) ]
\pm u [ \varphi (x \mp 2\xi ) - \varphi (x) ]
\eea
We compare with the action of the generator $S_{\xi,12}^0$ as  a difference
operator on functions of $x_{12} = x$ (\ref{s0phi}) and use
the notation $\tilde S_{\pm\xi,x}^0 \varphi (x)$ introduced there,
\be
\label{KLRS0}
2 \xi K_{L/R}^{12} (u) \varphi (x) \ = \  (\tilde S_{\mp \xi, x}^0 \ \pm \ u )
\ \  \left (\varphi (x \mp 2 \xi) - \varphi (x) \right ).
\ee
We know that $\varphi_{12,n}^{(\pm\xi), (0)} (x)$ are eigenfunctions
of $\tilde
S_{\pm \xi 12}^0 $ with eigenvalue $\ell_1 + \ell_2 + n$. We use also the
iterative relation (\ref{phinrec}) to obtain
\be
\label{KLRphin}
2 \xi K_{L/R}^{12} (u)  \ \ \varphi_{12,n}^{(\mp\xi), (0)} (x) =
\mp 2 n \
(\ell_1 +\ell_2 + n  \pm u) \ \ \varphi_{12,n}^{(\mp\xi), (0)} (x).
\ee

We use now the latter result and the eigenvalue relations (\ref{Reigenv})
to evaluate both sides of (\ref{RKphi}) and obtain
\be
\label{rhophi}
-(\ell_1 + \ell_2 + n + u) \rho_{\ell_1,  \ell_2,  n - 1} \
\varphi_{12,n-1}^{(-\xi), (0)} (x)  =
(\ell_1 + \ell_2 + n - u)  \rho_{\ell_1,  \ell_2,  n }
\varphi_{12,n-1}^{(-\xi), (0)} (x).
\ee
This proves the compatibility of the eigenvalue relations with the defining
conditions of the universal R operator and results in the
iterative relation for the eigenvalues with the solution
\be
\label{rhon}
\rho_{\ell_1,  \ell_2,  n} = const (-1)^n
{ \Gamma (\ell_1 + \ell_2 + n + u + 1) \Gamma (\ell_1 + \ell_2 -u) \over
\Gamma (\ell_1 + \ell_2 +   u ) \Gamma (\ell_1 + \ell_2 +n +1 -u) }.
\ee
The eigenvalues do not depend on the deformation parameter $\xi$
and coincide with the undeformed ones.
This is expected from the twist relation between the deformed and undeformed
universal R operators to be considered now.
We compare the eigenvalue relations (\ref{Reigenv}) with the one for the
undeformed universal R operator,
\be
\label{R0eigenv}
R_{12,\ell_1,\ell_2}^{(0)}(u) \ \ \phi_{12, n}^{(m)}  =
\rho^0_{\ell_1,  \ell_2, n} \ \ \ \phi_{12, n}^{(m)}.
\ee
We specify the twist relation (\ref{FRF}) to the considered
generic representations $V_1^{\ell_1} \otimes V_2^{\ell_2}$,
\be
\label{FRFell}
R_{12,\ell_1,\ell_2}^{(\xi)}(u)  \ = \
F_{\xi, 21}^{\ell_2,\ell_1} \ \
R_{12,\ell_1,\ell_2}^{(0)}(u)  \ \ (F_{\xi, 12}^{\ell_1, \ell_2} ) ^{-1}
\ee
writing the twist element  (\ref{Fsigma}) in the appropriate representations
explicitely as operators on functions of $x_1,
x_2$,
\be
\label{F12x}
F_{\xi,12}^{\ell_1,\ell_2} = \exp (S_{0,1}^0 \ \sigma_{\xi,2} ) =
\exp (\xi S_{0,x_1}^0 \ \dd_{x_2}  ),
\ee
where
$$ \sigma_{\xi,2} = - \ln ( 1 - 2 \xi S_{0,x_2}^-) = \xi \dd_{x_2}, $$
$$ S_{0,x_1}^0 = \frac{1}{2} \left ( X_1^{(-\xi, 2\ell_1 -1)}
D_{x_1}^{(-\xi)} +   D_{x_1}^{(-\xi)} X_1^{(-\xi, 2\ell_1 -1)}
\right ) + \ell_1 - \frac{1}{2}. $$
The comparison of the eigenvalue relations (\ref{Reigenv}, \ref{R0eigenv})
with repect to the twist relation
implies the non-deformation of the eigenvalues, $\rho_{\ell_1,  \ell_2,  n}
= \rho^0_{\ell_1,  \ell_2,  n}$, and the following relation of the basis
polynomials of the deformed and induced undeformed representations,
\be
\label{phiFphi}
\phi_{12, n}^{(m)} (x_1,x_2) \ = \ (F_{\xi 12}^{\ell_1,\ell_2})^{-1}
 \varphi_{12, n}^{(\xi),(m)} (x_1,x_2) =
: \varphi_{12, n}^{(\xi),(m)} (x_1,x_2 - S_{0,x_1}^0) : \cdot 1
\ee
The last expression is to be understood from the expansion of the
polynomial
in powers of $x_1$ and $x_2$, ordering in each term the operators
$ S_{0,x_1}^0 $ to the left from any power of $x_1$ and acting with the
resulting operator on the constant $1$.

Above we have given an explicite expression for the polynomials representing
the highest weights in the deformed tensor product (\ref{phin0x12})
and a rule for the computation of the basis polynomials of the undeformed
tensor product in the induced picture (\ref{phixnmrule}).
The relation between these series of polynomials is quite non-trivial; a
direct check has been done for some lowest order polynomials only.

\subsection{The integral form  at small $\xi$}

In the polynomial representation considered the universal R operator acts on
functions of two variables. We would like to represent this action in
integral form,
\be
\label{Rint}
R_{12,\ell_1,\ell_2}^{(\xi)}(u) \ \varphi(x_1,x_2) \ = \
\int dx_{1\p} dx_{2\p} \mathcal{R}^{(\xi)}_{\ell_1,\ell_2} (u; 
x_1,x_2; x_{1\p},x_{2\p}) \  \varphi(x_{1\p},x_{2\p}).
\ee
We assume that both integrations go along closed contours in order to have
simple integrations by part.

QYBE in the representastion of generic $\ell_1, \ell_2$ and $\ell_3 =
-\frac{1}{2}$ which serves as the defining conditions for the universal R
operator with the known Lax operators implies now conditions on the kernel
$\mathcal{R}$. We specify to the form with symmetry in $\xi \leftrightarrow
-\xi$ and write down the Lax matrix (\ref{Lxisym})
\bea
\label{Laxsigma}
L_{i,\ell_i}^{(\xi) sym } (u) = L_{i,\ell_i}
^{(\xi) sym} + u ( \ch(\xi \dd_i) I + (u+1)
\hat\sigma^- \sh(\xi \dd_i ) ),  \cr
L_{i, \ell_i} ^{(\xi) sym} = \frac{1}{2} \ch(\xi S_{\xi,i,\ell_i}^-) I +
\left ( \matrix { S_{\xi,i,\ell_i}^0 \ &\ \frac{1}{\xi} \sh(\xi S_{\xi,i,\ell_i}^-) \cr
S_{\xi,i,\ell_i}^+ \ &\ - S_{\xi, i}^0 \cr } \right ),
\eea
and also its transposition, appearing by partial integration with respect to
$x_i$,
\be
\label{LaxT}
(L_i^{(\xi) sym } (u))^T = - L_{i,\ell_i^T}^{(\xi) sym} +
(u+1)  ( \ch(\xi \dd_i) I - u \hat \sigma^- \sh(\xi \dd_i ) ),
\ee
Notice that $T$ does not transpose the $2\times 2$ matrix and that
 $\ell_i^T = 1-\ell_i $. We keep in mind that the generators in
$L_{i,\ell_i} $ are operators acting on $x_i$ in the
 representation $\ell_i$. We suppress the representation
label adopting the convention $\ell_{i\p} = 1- \ell_i, 1= 1,2$.
In these notations the defining condition for the kernel reads,
\bea
\label{L1L2kernel}
\left[ [L_2^{(\xi) sym} + v ( \ch(\xi \dd_2) I + (v+1)
\hat \sigma^- \sh(\xi \dd_2 ) )]  \right. \cr
[L_1^{(\xi) sym} + (u+v) ( \ch(\xi \dd_1) I + (u+v+1)
\hat\sigma^- \sh(\xi \dd_1 ) ] \cr
-
[L_{1\p}^{(\xi) sym} - (u+v+1) ( \ch(\xi \dd_{1\p}) I - (u+v)
\hat\sigma^- \sh(\xi \dd_{1\p} ) ]\cr\left.
  [L_{2\p}^{(\xi) sym} - (v+1) ( \ch(\xi \dd_{2\p}) I - v
\hat\sigma^- \sh(\xi \dd_{2\p} ) ]  \right ] \cr
\mathcal{R}^{(\xi)}_{\ell_1,\ell_2} (u;x_1,x_2; x_{1\p},x_{2\p})
= 0
\eea
From this matrix condition we obtain two conditions  projecting
$$ \left ( \matrix {\hat A \mathcal{R} & \hat B \mathcal{R} \cr
\hat C \mathcal{R} & \hat D \mathcal{R} \cr}
\right ) = 0
$$
by multiplication from the left by the row $(x_l, 1)$ and from the right
by the column $(1, - x_r)^T$. In the first case we put $x_l = x_2$ and
$x_r = x_{2\p}$, in the second case $x_l = x_1$ and $x_r = x_{1\p}$ after
the differential operations are done, i.e. we understand that the operators
$\hat A, \hat B, \hat C, \hat D$ act on $\mathcal{R}$ only. In this way we
obtain the defining relations for the kernel of the universal R operator
\be
\label{ABCDR}
\left ( x_2 \hat A + \hat C - x_2 x_{2\p} \hat B + x_{2\p} \hat D \right ) \
\mathcal{R}^{(\xi)}_{\ell_1,\ell_2} (u;x_1,x_2; x_{1\p},x_{2\p})
= 0,
\ee
$$
\left ( x_{1\p} \hat A + \hat C - x_{1\p} x_{1} \hat B + x_{1} \hat D \right ) \
\mathcal{R}^{(\xi)}_{\ell_1,\ell_2} (u;x_1,x_2; x_{1\p},x_{2\p})
= 0.
$$
The operator matrix in (\ref{L1L2kernel}) is symmetric under the simultaneous
exchange
\be
\label{crossing}
1 \leftrightarrow {2\p}, \ \ \ \  2 \leftrightarrow {1\p} , \ \ \ \
v \leftrightarrow -1-u-v
\ee
and by this crossing symmetry the latter two conditions are
transformed into each other.

We are going to solve the conditions (\ref{ABCDR}) perturbatively in $\xi$
up to the second order. There is no first order contribution,
\be
\label{xiexpan}
\mathcal{R}^{(\xi) sym (\ell_1,\ell_2)} =
\mathcal{R}^{[0] {\ell_1,\ell_2}} + \xi^2
\mathcal{R}^{[2] {\ell_1,\ell_2}} + {\cal O} (\xi^4),
\ee
since $\hat A, \hat B,\hat C, \hat D$ are symmetric in $\xi$.
The kernel of the original non-symmetric operator is calculated from the
symmetric one by
\be
\label{nonsym}
\mathcal{R}^{(\xi) (\ell_1,\ell_2)} (x_1,x_2; x_{1\p},x_{2\p})
\ = \ e^{- u \xi (\dd_1 + \dd_{1\p})} \
\mathcal{R}^{(\xi)sym(\ell_1,\ell_2)} (x_1,x_2; x_{1\p},x_{2\p}).
\ee
The first condition (\ref{ABCDR}) at order $\xi^0$ reads
\bea
\label{ABCD0}
(v+{\frac 12} - \ell_2) [ x_{2\p 1} x_{12} \dd_1 - (u+v+{\frac 12} + \ell_1)
x_{12} - (u+v+{\frac 12} - \ell_{1} ) x_{2\p 1} ]
\mathcal{R}^{[0]} \cr
+
(v-{\frac 12} + \ell_2) [ x_{ 1\p 2} x_{1\p 2\p} \dd_{1\p} -
(u+v+{\frac 32} - \ell_1)
x_{1\p2\p} - (u+v-{\frac 12} + \ell_{1} ) x_{2 1\p} ]
\mathcal{R}^{[0]}  = 0.
\eea
$v$ appears here as a free parameter. The implication for $v= \ell_2 -{\frac
12}$ is
\be
\dd_1 \left ( { x_{2\p 1}^{u+1+\ell_1 -\ell_2} \over
x_{12}^{u+1-\ell_1-\ell_2 } }   \right )
\mathcal{R}^{[0]} = 0
\ee
and for $v= {\frac 12} - \ell_2$
\be
\dd_{1\p} \left ( { x_{2 1\p}^{u+1+\ell_2 -  \ell_1} \over
x_{1\p 2\p}^{u-1+\ell_1+\ell_2 } }   \right )
\mathcal{R}^{[0]}= 0
\ee
The two implications can be expressed together as
\be
\label{R0cond}
\dd_a \left ( { x_{2\p 1}^{u+1+\ell_1 -\ell_2}
x_{2 1\p}^{u+1+\ell_2 -  \ell_1} \over
x_{12}^{u+1-\ell_1-\ell_2 }  x_{1\p 2\p}^{u-1+\ell_1+\ell_2 } }
 \right ) \mathcal{R}^{[0]} = 0
\ee
for $a= 1, 1\p$ The second condition in (\ref{ABCDR}) implies in
order $\xi^0$ the analogous condition for $a=2, 2\p$ which is clear from the
mentioned crossing symmetry (\ref{crossing}).
In this way we recover the result for the kernel of the undeformed $sl(s)$
symmetric universal R operator.
\be
\label{ndk}
\mathcal{R}^{[0]}(u)=c(u)\frac{x_{12}^{u-\ell_1-\ell_2+1}x_{1\p
2\p}^{u+\ell_1+\ell_2-1}}
{x_{2\p1}^{u+1+\ell_1-\ell_2}x_{21\p}^{u+1-\ell_1+\ell_2}}.
\ee
The first condition in (\ref{ABCDR}) at
 order $\xi^2$ reads
$$
(v+\frac{1}{2} -\ell_2)[x_{2\p1}x_{12}\dd_1-(u+v+\frac{1}{2} +\ell_1)x_{12}-
(u+v+\frac{1}{2} -\ell_1)x_{2\p1}]
\mathcal{R}^{[2]}(u)+$$
$$+(v+\ell_2 - \frac{1}{2})
[x_{1\p 2\p}x_{21\p}\dd_{1\p}-(u+v+\frac{3}{2} + \ell_1)x_{1\p 2\p}-
(u+v+\frac{1}{2} - \ell_1)x_{21\p}] \mathcal{R}^{[2]}(u)+
$$
$$
+(v+\frac{1}{2} -\ell_2)[\frac 16x_{12}\dd_1^3x_{2\p1}-
\frac 12(u+v)\dd_1^2(x_{12}+x_{2\p1})
+(u+v)(u+v+1)\dd_1] \mathcal{R}^{[0]}(u)+
$$
\be
\label{I2}
+(v+\ell_2 - \frac{1}{2})
[\frac
16x_{21\p}\dd_{1\p}^3x_{1\p 2\p}-\frac 12(u+v+1)\dd_{1\p}^2(x_{1\p
2\p}+x_{21\p})
+(u+v)(u+v+1)\dd_{1\p}]\mathcal{R}^{[0]}(u)+
\ee
$$
\frac v2\dd_2^2[x_{2\p1}x_{12}\dd_1-(u+v+\frac{1}{2} +\ell_1)x_{12}-
(u+v+\frac{1}{2} - \ell_1)x_{2\p1}]\mathcal{R}^{[0]}(u)+
$$
$$
+\frac v2\dd_2^2[x_{1\p2\p}x_{21\p}\dd_{1\p}-(u+v+\frac{3}{2}
-\ell_1)x_{1\p 2\p}- (u+v+\ell_1 - \frac{1}{2})x_{21\p}]
\mathcal{R}^{[0]}(u)+
$$
$$
+v^2\dd_2[u+v+\frac{1}{2} +\ell_1-x_{2\p1}\dd_1]R^{[0]}(u)+
 v^2\dd_{2\p}[u+v+\frac{3}{2} -\ell_1 -x_{21\p}\dd_{1\p}]
\mathcal{R}^{[0]}(u)=0.
$$
The second condition in (\ref{ABCDR}) leads to the analogous implication
obtained from the latter by symmetry (\ref{crossing}). By choosing
$v= \pm (\frac{1}{2} - \ell_2)$ and using the result for $R^{[0]}(u)$
we obtain four differential equations for the  function $R^{(2)}(u)$,
which can be written together in the form
$$
\dd_a[{\mathcal{R}^{[2]}(u)}/{\mathcal{R}^{[0]}(u)}]=
$$
$$
=\!\dd_a\!\left[\!\frac 1{12}(u\!+\!1\!+\!\ell_1\!-\!\ell_2)
(u\!+\!2\!+\!\ell_1\!-\!\ell_2)(2u\!-\!\ell_1\!+\!\ell_2)\frac
1{x_{2\p1}^2}\!+ \right. $$
$$
\frac {2\ell_1 -1}4(u\!+\!\ell_1\!+\!\ell_2 -1)(u\!+\!1\!-\!\ell_1\!+\!\ell_2)
\frac 1{x_{21\p}x_{1\p 2\p}}\!-
$$
\be
\label{R2a}
\!-\frac 1{12}(u\!-\!\ell_1\!-\!\ell_2 +1)(u\!\!-\!\ell_1\!-\!\ell_2)
(2u\!+1\!+\!\ell_1\!+\!\ell_2)\frac 1{x_{12}^2} \!-\!
\frac {2 \ell_1-1}4(u\! +1 -\!\ell_1\!-\!\ell_2)(u\!+\!1\!+\!\ell_1\!-\!\ell_2)
\frac 1{x_{12}x_{2\p1}}\!-
\ee
$$
\!-\frac 1{12}(u\!+1 -\!\ell_1\!-\!\ell_2)(u\!-\!\ell_1\!-\ell_2)
(2u\!+1\!+\!\ell_1\!+ \ell_2)\frac 1{x_{12}^2}\!-
\frac {2\ell_2-1}4(u\!-\!\ell_1\!-\!\ell_2 +1)(u\!+\!1\!-\!\ell_1\!+\!\ell_2)
\frac 1{x_{12}x_{21\p}}\!
$$
$$
-\!\frac 1{12}(u\!+\!1\!-\!\ell_1\!+\!\ell_2)(u\!+\!2\!-\!\ell_1\!+\!\ell_2)
(2u\!+\!\ell_1\!-\!\ell_2)\frac 1{x_{21\p}^2}\!+\frac {2\ell_2-1}4(u\!+
\!\ell_1\!+ \ell_2 -1)(u\!+\!1\!+\!\ell_1\!-\!\ell_2)
\frac 1{x_{2\p1}x_{1\p 2\p}}\!
$$
$$
\!-\frac 1{12}(u\!+\!\ell_1\!+\!\ell_2-1)(u\!-\!2\!+\!\ell_1\!+\!\ell_2)
(2u\!+3\!-\!\ell_1\!-\!\ell_2)\frac 1{x_{1\p2\p}^2} $$
$$ \left.
-\!\frac{2\ell_2-1}4(u\!-
\!\ell_1\!-\!\ell_2 +1)(u+\!1\!-\!\ell_1\!+ \ell_2)\frac 1{x_{12}x_{21\p}}
\right]\!,
$$
The index $a$ labels the derivatives with respect to the points
$x_1,x_2,x_{1\p}, x_{2\p}$.
The ratio of the second order correction to the undeformed kernel is equal
to the square bracket on the right hand side up to a constant. This constant
can be absorbed by an unessential $\xi^2$ correction to $c(u)$ in
(\ref{ndk}). As the result up to the order $\xi^2$
the integral kernel of $R^{sym}$ has form:
\bea
\label{Rsym2}
\mathcal{R}^{(\xi)sym(\ell_1,\ell_2)}(u)=e^{u\xi(\dd_1+\dd_{1\p})}
\mathcal{R}^{(\xi)
(\ell_1,\ell_2)}(u)= \cr
\mathcal{R}^{(0) (\ell_1,\ell_2)}(u)\left[1-\frac
{\xi^2}{12}[(2u+1+\ell_1+\ell_2)a_{11}-(2u+\ell_1-\ell_2)a_{22} \right. \cr
+(2u+3-\ell_1-\ell_2)a_{33}-(2u+\ell_2-\ell_1)a_{44}]  \cr
\left. -\frac{\xi^2}4
[(2\ell_1-1)(a_{14}-a_{23})+(2\ell_2-1)(a_{12}-a_{34})]
+ {\cal O}(\xi^2) \!\right]\!,
\eea
here we have introduced the notations
$$
a_{11}=\frac{(u+1 -\ell_1-\ell_2)(u-\ell_1-\ell_2)}{x_{12}^2},\quad
a_{22}=\frac{(u+1+\ell_2-\ell_1)(u+2+\ell_2-\ell_1)}{x_{21\p}^2},
$$
$$
a_{33}=\frac{(u+\ell_1+\ell_2-1)(u-2+\ell_1+\ell_2)}{x_{1\p2\p}^2},\quad
a_{44}=\frac{(u+1+\ell_1-\ell_2)(u+2+\ell_1-\ell_2)}{x_{2\p1}^2},
$$
$$
a_{12}=\frac{(u+1 -\ell_1-\ell_2)(u+1+\ell_2-\ell_1)}{x_{12}x_{21\p}},\quad
a_{13}=\frac{(u+1-\ell_1-\ell_2)(u-1+\ell_1+\ell_2)}{x_{12}x_{1\p 2\p}},
$$
$$
a_{14}=\frac{(u+1 -\ell_1-\ell_2)(u+1+\ell_1-\ell_2)}{x_{12}x_{2\p1}},\quad
a_{23}=\frac{(u+1+\ell_2-\ell_1)(u+\ell_1+\ell_2-1)}{x_{21\p}x_{1\p 2\p}},
$$
$$
a_{24}=\frac{(u+1+\ell_2-\ell_1)(u+1+\ell_1-\ell_2)}{x_{21\p}x_{2\p1}},\quad
a_{34}=\frac{(u-1 +\ell_1+\ell_2)(u+1+\ell_1-\ell_2)}{x_{1\p 2\p}x_{2\p 1}}.
$$

\section{Summary}

Our study relies on the known algebraic structure of the $\xi$ deformation
of the $sl(2)$ Yangian as reviewed in Sect. 2. 
We have obtained the known fundamental representation $R$ operator and the
Lax operator both from the symmetry condition and from the Drinfeld twist
transformation. We have written the Lax operator in two forms, one in terms
of the deformed generators and one in terms of the undeformed generators. 
Therefrom we have established the relation of between the deformed and
undeformed generators. 

The defining condition of the universal $R$ operator
following from QYBE in the spirit of \cite{TTF} has been formulated. 

 The structure of representations by polynomials in one
variable and of the tensor product representations by polynomials in two
variables has been investigated. 
We have written the deformed and undeformed generators in terms
of several {\sl Heisenberg} canonical pairs, the original one $(y, \dd )$
and further ones expressed as non-trivial tranformations of the former.
The particular form of deformed generators, where $S^-$ is the 
infinitesimal translation $\dd $, is preferred in view of possible physical
applications where momentum conservation is important.  Then the twist
relations between the deformed and undeformed generators induce a quite
involved expression of the latter in terms of the original pair $(x, \dd )$. 

The basis polynomials of the deformed representation and
also the undeformed representation polynomials in the non-trivial induced
form have been constructed. 
The resulting expressions involve products of the variable $x$  
displaced  by subsequent units of the deformation paramenter $\xi$. 
 Also the polynomials of $x_1, x_2$ representing the lowest weight
states of the irreducible representations in the tensor products have been
calculated. They have been
expressed in terms of products of the difference $x_{12}$  with
displacements depending on the representations $\ell_1, \ell_2$ and 
increasing in units of $\xi$ and also summed up into a generating function.

With the explicite lowest weight polynomials we have studied 
the defining condition of the universal $R$ operator writing the operators
involved in terms of $(x, \dd )$. The resulting recurrence relation confirms 
the known fact that the eigenvalues of the $R$ operator coincide with the
ones of the undeformed $R$ operator, as expected from the relation between
them by the twist transformation. This transformation implies also a
non-trivial relation between the series of polynomials in two variables 
representing the deformed tensor products and the induced form of the
undeformed tensor products.   

 Finally the universal $R$ operator in integral form has been considered.
The defining conditions on the integral kernel
 have been solved perturbatively in $\xi$ up to the first
non-trivial order $\xi^2$ applying the projection method of 
\cite{Derkachov:2001sx}.

We have treated the Jordanian deformation of the $sl(2)$ bi-algebra
structure in view of physical applications where the undeformed $sl(2)$
represents infinitesimal conformal tranformations with the dimensions of the
representation being in general infinite. We have worked out these
representations in terms of polynomial (wave) functions, generators and Lax
operators in terms of differential operators and the universal $R$ operator
in spectral and integral forms in close analogy to the previously studied
cases of supersymmetric extension $(sl(2|1) )$ 
or standard $q$ deformation of the $sl(2)$ bi-algebra structure.
It will be interesting to study applications to integrable systems with
conformal symmetry broken in such a way that translation symmetry is
preserved.

\section*{Acknowledgements}
The collaboration of groups of Leipzig University and Yerevan Physics
Institute has been supported by German Federal Ministry BMBF. 
One of us (S.D.) is grateful to Saxonian Ministry SMWK and to NTZ for support
and one of us (D.K.) benefitted also from a support by Volkswagen Stiftung.

\section{Appendices}
\setcounter{equation}{0}
\subsection{Appendix A}

We show that the deformed commutation relations follow from the
relations (\ref{SxiS0}) of the deformed and undeformed generators.
The following commutators with functions of $S_0^-$ are useful,
\be
\label{S0f}
[S_0^0, f(S_0^-) ] = - \alpha \dd_{\alpha} f(\alpha) |_{\alpha = S_0^-},
\ee
\be
\label{S+f}
[S_0^+, f(S_0^-) ] = 2 \dd_{\alpha} f(\alpha ) \ S_0^0 - \alpha
\dd^2_{\alpha} f(\alpha) |_{\alpha = S_0^-},
\ee

\be
\label{SS}
[S_0^0 (S_0^0 + \frac{1}{2}), f(S_0^-) ] =  - 2 \alpha    \dd_{\alpha}
f(\alpha ) \ S_0^0 + (\alpha^2 \dd^2_{\alpha}  + \frac{1}{2} \alpha
\dd_{\alpha})  f(\alpha) |_{\alpha = S_0^-},
\ee
They imply in particular,
\be
\label{S+sqrt}
[S_0^+ + 2 \xi S_0^0 (S_0^0 + \frac{1}{2}) ]
\sqrt{1- 2 \xi S_0^-} =
\sqrt{1- 2 \xi S_0^-}
[S_0^+ + 2 \xi S_0^0 (S_0^0 - \frac{1}{2}) ]
\ee
\be
\label{S0sqrt}
\left [S_0^0, \sqrt{1-2\xi S_0^-} \right ] \ = \ 
{\xi S_0^- \over \sqrt{1- 2\xi S_0^-} }
\ee

Let us check first the relation
\be
[S_{\xi}^0, S_{\xi}^- ] =  - \frac{1}{\xi} \sh(\xi S_{\xi}^- )
\ee
We substitute the expessions in terms of the undeformed generators
(\ref{SxiS0}). The right-hand side is equal to
$- S_0^- (1- 2\xi S_0^-)^{-\frac{1}{2}}$. The commutator on the
left-hand side is calculated by using (\ref{S0f}) and the result is
easily seen to coincide with the one for the right-hand side.

Let us check now the relation
\be
[S_{\xi}^+, S_{\xi}^- ] =   \frac{1}{2} \{\ch(\xi S_{\xi}^-),
S_{\xi}^+\} \ee
We write the right-hand side in terms of the undeformed generators and
apply (\ref{S+sqrt}),
$$ \frac{1}{4} \left ( {1\over
 \sqrt{1- 2\xi S_0^-} } +
\sqrt{1- 2\xi S_0^-}  \right ) \ 
[S_0^+ + 2 \xi S_0^0 (S_0^0 + \frac{1}{2}) ] \ 
\sqrt{1- 2 \xi S_0^-} + $$ 
$$  \frac{1}{4}
[S_0^+ + 2 \xi S_0^0 (S_0^0 + \frac{1}{2}) ] \ 
\sqrt{1- 2 \xi S_0^-} \ 
 \left ({1\over
 \sqrt{1- 2\xi S_0^-} } +
\sqrt{1- 2\xi S_0^-}  \right )  = $$
$$
{1 - \xi S_0^- \over 2}
[S_0^+ + 2 \xi S_0^0 (S_0^0 - \frac{1}{2}) ] +
[S_0^+ + 2 \xi S_0^0 (S_0^0 + \frac{1}{2}) ]
{1 - \xi S_0^- \over 2} = $$
$$ S_0^+ + \xi \left ( 2 (S_0^{0})^2 - \frac{1}{2} (S_0^+ S_0^- + S_0^-
S_0^+) \right ) - \xi^2 \left ( S_0^- (S_0^0)^2  + (S_0^0)^2 S_0^- - 
\frac{1}{2} S_0^- \right ). $$
We write also the left-hand side in terms of the undeformed generators
and apply (\ref{S+sqrt}, \ref{S0sqrt}),
$$-\frac{1}{2} \xi S_0^-
[S_0^+ + 2 \xi S_0^0 (S_0^0 - \frac{1}{2}) ] +
S_0^0 (1-2\xi S_0^-)
[S_0^+ + 2 \xi S_0^0 (S_0^0 - \frac{1}{2}) ] + $$ $$
[S_0^+ + 2 \xi S_0^0 (S_0^0 + \frac{1}{2}) ]
\frac{1}{2} \xi S_0^- -
[S_0^+ + 2 \xi S_0^0 (S_0^0 + \frac{1}{2}) ]
(S_0^0 + \xi S_0^- (1- 2 S_0^0))   =                $$
 $$ S_0^0 S_0^+ - S_0^+ S_0^0 +   \xi [ -\frac{1}{2} S_0^- S_0^+ + 2
(S_0^0)^2 (S_0^0- \frac{1}{2} ) - 2 S_0^0 S_0^- S_0^+ + \frac{1}{2}
S_0^+ S_0^- - 2 (S_0^0)^2 (S_0^0 + \frac{1}{2}) ) - S_0^+ S_0^-
(1-2S_0^0) ] + $$ $$
\xi^2 [ -S_0^- S_0^0 (S_0^0 -\frac{1}{2} ) - 4 S_0^0 S_0^- S_0^0 (S_0^0
- \frac{1}{2}) + S_0^0 (S_0^0 + \frac{1}{2}) S_0^- -
S_0^0 (1+ 2 S_0^0 ) S_0^- (1-2S_0^0) ] $$
Now it is easy to check that the coefficients of $\xi^2, \xi^1, \xi^0$
of both sides are equal.
The right hand side of the third commutator is equal to
$$ - {\xi S_0^- \over \sqrt{1-2\xi S_0^-} } + 2 S_0^0
 \sqrt{1-2\xi S_0^-} $$
The commutator on the left-hand side is
$$ - \frac{1}{2\xi}
[S_0^+ + 2 \xi S_0^0 (S_0^0 + \frac{1}{2}), \ln (1-2\xi S_0^-) ]
\sqrt {1-2\xi S_0^-} $$
We apply (\ref{S+f}, \ref{SS}). Then it is easy to see that the result
is equal to the right hand side.

\subsection{Appendix B}

We list the operator valued matrix elements of $K_L(u)$ and $K_R (u)$
introduced in (\ref{ybeu}).

$$
\xi K^{11}_L=(\frac 12\ch\xi S_{\xi,1}^-+S_{\xi,1}^0)
(\frac 12 \ch\xi S_{\xi,2}^-+S_{\xi,2}^0)
+\frac 1\xi\sh\xi S_{\xi,1}^-S_{\xi,2}^++
$$
$$
+{\frac u2}\left(e^{-\xi S_{\xi,1}^-}(\frac 12
\ch\xi S_{\xi,2}^-+S_{\xi,2}^0)-(\frac 12\ch\xi S_{\xi,1}^-+S_{\xi,1}^0)
e^{-\xi S_{\xi,2}^-}+
\sh\xi S_{\xi,1}^-(2S_{\xi,2}^0+\sh\xi S_{\xi,2}^-)\right),
$$
$$
 \xi K^{12}_L= \frac u2\left(e^{-\xi S_{\xi,1}^-}\sh\xi
S_{\xi, 2}^--\sh\xi S_{\xi,1}^-e^{\xi S_{\xi,2}^-}\right)+
(\frac 12\ch\xi S_{\xi,1}^-+S_{\xi,1}^0)\sh\xi S_{\xi,2}^-+
\sh\xi S_{\xi,1}^-(\frac 12\ch\xi S_{\xi,2}^--S_{\xi,2}^0),
$$
$$
\xi K^{22}_L=(\frac 12\ch\xi S_{\xi,1}^--S_{\xi,1}^0)(\frac 12 \ch\xi
S_{\xi,2}^--S_{\xi,2}^0) +\frac 1\xi\sh\xi S_{\xi,2}^-S_{\xi,1}^++
$$
$$
+\frac u2\left(e^{\xi S_{\xi,1}^-}(\frac 12
\ch\xi S_{\xi,2}^--S_{\xi,2}^0)-(\frac 12\ch\xi
S_{\xi,1}^-+S_{\xi,}1^0)e^{\xi
S_{\xi,2}^-}+ \sh\xi S_{\xi,2}^-(2S_{\xi,1}^0+\sh\xi
S_{\xi,1}^-)\right), $$
$$
\xi K^{21}_L=S_{\xi,1}^+(\frac 12\ch\xi S_{\xi,2}^-+S_{\xi,2}^0)+(\frac
12 \ch\xi S_{\xi,1}^--S_{\xi,1}^0) S_{\xi,2}^++
$$
$$
\!\!+\!\frac u2\left(\!\xi(2S_{\xi,1}^0+\sh\xi S_{\xi,1}^-)\!(\frac
12\ch\xi S_{\xi,2}^-+S_{\xi,2}^0)\!+\!
e^{\xi S_{\xi,1}^-}S_{\xi,2}^+-e^{-\xi
S_{\xi,2}^-}S_{\xi,1}^+\!-\!\xi(\frac
12\ch\xi S_{\xi,1}^--S_{\xi,1}^0) \!(2S_{\xi,2}^0+\sh\xi
S_{\xi,2}^-)\!\right) $$
The expressions for the elements of right matrix are  similar
$$
\xi K^{11}_R=(\frac 12\ch\xi S_{\xi,1}^-+S_{\xi,1}^0)(\frac 12 \ch\xi
S_{\xi,2}^-+S_{\xi,2}^0) +\frac 1\xi\sh\xi S_{\xi,2}^-S_{\xi,1}^++
$$
$$
+\frac u2\left(e^{-\xi S_{\xi,1}^-}(\frac 12
\ch\xi S_{\xi,2}^-+S_{\xi,2}^0)-(\frac 12\ch\xi
S_{\xi,1}^-+S_{\xi,1}^0)e^{-\xi
S_{\xi,2}^-}+ \sh\xi S_{\xi,2}^-(2S_{\xi,1}^0+\sh\xi
S_{\xi,1}^-)\right), $$
$$
\xi K^{12}_R=\frac u2\left(e^{\xi S_{\xi,1}^-}\sh\xi S_{\xi,2}^--\sh\xi
S_{\xi,1}^-e^{-\xi S_{\xi,2}^-}\right)+
(\frac 12\ch\xi S_{\xi,1}^--S_{\xi,1}^0)\sh\xi S_{\xi,2}^-+
\sh\xi S_{\xi,1}^-(\frac 12\ch\xi S_{\xi,2}^-+S_{\xi,2}^0),
$$
$$
\xi K^{22}_R=(\frac 12\ch\xi S_{\xi,1}^--S_{\xi,1}^0)(\frac 12 \ch\xi
S_{\xi,2}^--S_{\xi,2}^0) +\frac 1\xi\sh\xi S_{\xi,1}^-S_{\xi,2}^++
$$
$$
+\frac u2\left(e^{\xi S_{\xi,1}^-}(\frac 12
\ch\xi S_{\xi,2}^--S_{\xi,2}^0)-(\frac 12\ch\xi
S_{\xi,1}^-+S_{\xi,1}^0)e^{\xi
S_{\xi,2}^-}+ \sh\xi S_{\xi,1}^-(2S_{\xi,2}^0+\sh\xi
S_{\xi,2}^-)\right), $$
$$
\xi K^{21}_R=S_{\xi,1}^+(\frac 12\ch\xi S_{\xi,2}^--S_{\xi,2}^0)+(\frac
12 \ch\xi S_{\xi,1}^-+S_{\xi,1}^0) S_{\xi,2}^++
$$
$$
\!\!+\!\frac u2\left(\!\xi(2S_{\xi,1}^0+\sh\xi S_{\xi,1}^-)\!(\frac
12\ch\xi S_{\xi,2}^--S_{\xi,2}^0)\!
+\!e^{-\xi S_{\xi,1}^-}S_{\xi,2}^+\!-\!
e^{\xi S_{\xi,2}^-}S_{\xi,1}^+-\xi(\frac 12\ch\xi
S_{\xi,1}^-+S_{\xi,1}^0)\! (2S_{\xi,2}^0+\sh\xi S_{\xi,2}^-)\!\right)
$$

\subsection{Appendix C}

The eigenvalue condition (\ref{SJphi}) can be written as
\be
\label{SJphi1}
\frac{1}{2} (x \dd^{\xi}_x + x\dd_x^{\xi} ) \varphi^{(m)} (x) = 
(m+\frac{1}{2} ) \varphi^{(m)} (x) 
\ee
It is actually independent of $\ell$. We calculate the action of the
opertor on l.h.s. on products,
\bea
\frac{1}{2} (x \dd^{\xi}_x + x \dd_x^{\xi} )
\prod_{k_=-m_1}^{m_1} (x+ k\xi) = 
(2m_1+\frac{3}{2} ) \prod_{k_=-m_1}^{m_1} (x+ k\xi) \cr
+ \xi^2 2 m_1 (m_1 + \frac{1}{2})^2 
\prod_{k_=-m_1+1}^{m_1-1} (x+ k\xi), \cr
\frac{1}{2} (x \dd^{\xi}_x + x \dd_x^{\xi} )
x\prod_{k_=-m_1}^{m_1} (x+ k\xi) = 
(2m_1+\frac{5}{2} ) x \prod_{k_=-m_1}^{m_1} (x+ k\xi) \cr
+ \xi^2 2 (m_1+1) (m_1 + \frac{1}{2})^2 
x \prod_{k_=-m_1+1}^{m_1-1} (x+ k\xi).
\eea
From these relation we understand that the polynomial eigenfunctions obeying
(\ref{SJphi1}) are finite sums of such products,
\bea
\varphi^{(2m_1+1)} (x) = \sum_{k=0}^{m_1} a_k^{(m_1)} \xi^{2k} 
\prod_{k_1 =-m_1+k}^{m_1-k} (x+ k_1\xi), \cr
\varphi^{(2m_1+2)} (x) = \sum_{k=0}^{m_1+1} b_k^{(m_1)} \xi^{2k} 
 \ x \prod_{k_1 =-m_1+k}^{m_1-k} (x+ k_1\xi).
\eea
Substituting this into (\ref{SJphi1}) leads to simple iterative relations for
the coefficients with the solution
\bea
a_k^{(m_1)}  = \left (\matrix {m_1 \cr k }\right) 
\ \left ( { (2m_1 +1)!! \over (2(m_1-k) +1)!! \ 2^{k} }
\right )^2, \cr
b_k^{(m_1)}  = \left (\matrix {m_1 +1 \cr  k } \right) 
\ \left ({ (2m_1 +1)!! \over (2(m_1-k) +1)!! \ 2^{k} } \right )^2.
\eea

To derive the generating function $G_1(x, t)$ (\ref{repf})
we write $\varphi_{\ell}^{(m)}$ as a {\sl Fourier} integral and
$\varphi_{\ell}^{(m)} = (X_J)^m  \varphi_{\ell}^{(0)}$
as
\be
\label{varpin}
\varphi^{(m)}(x)
=\frac 1{2^m}\left(x(1+\ch\xi\dd)+\frac\xi 2\sh\xi\dd\right)^m
\int dka_0(k)e^{ikx}.
\ee
 The
Fourier image $a_0(k)$ of $\varphi_0(x)=1$ is given by delta-function
$\delta(k)$.
Now it is convenient to do the change of variables $\a=\tan(k\xi/2)$,
$$
\left(x(1+\cos k\xi)\frac d{dk}-\frac 12\sin k\xi\right)^n=
(\cos (k\xi/2))^{-1}(\dd/\dd\a)^n\cos (k\xi/2).
$$
Then the series of the generating function
can be summed up using {\sl Taylor}'s formula
$$
G_1(x,t)=\int dk\frac{a_0(k)}{\cos k\xi/2}\sum_{n=0}^\infty\frac 1{n!}
\left(\frac{t\xi}{2i}\frac\dd{\dd\a}\right)^ne^{ikx}\cos k\xi/2=
$$
\be
\label{repfunk}
=\int dka_0(k)\sqrt{1+\a^2(k)}\left(\frac{1+i(\a+\frac{t\xi}{2i}}
{1-i(\a+\frac{t\xi}{2i}}\right)^{\frac x\xi}
\left({1+i(\a+\frac{t\xi}{2i}}\right)^{-\frac 12}=
\ee
$$
\left(\frac{1+\frac{t\xi}2}{1-\frac{t\xi}2}\right)^{\frac x\xi}
\left(1+\frac{t\xi}2\right)^{-\frac 12}
\left(1-\frac{t\xi}2\right)^{-\frac 12}=
\frac{\left(1+\frac{t\xi}2\right)^{\frac x\xi-\frac 12}}
{\left(1-\frac{t\xi}2\right)^{\frac x\xi+\frac 12}}.
$$


\begin{thebibliography}{99}


\bibitem{LevPadua} L.N. Lipatov, {\it High-energy asymptotics of multicolor
QCD and exactly solvable lattice models}, Padova preprint DFPD-93-TH-70B;
and
\newline
JETP Lett. B342 (1994)596.


\bibitem{FK}
L.~D.~Faddeev and G.~P.~Korchemsky,
Phys.\ Lett.\ B {\bf 342} (1995) 311
[arXiv:hep-th/9404173].




\bibitem{DeVega:2001pu}
H.~J.~De Vega and L.~N.~Lipatov,
Phys. \ Rev. \ D{\bf 64} (2001) 114019,
arXiv:hep-ph/0107225.






\bibitem{Derkachov:2001yn}
S.~E.~Derkachov, G.~P.~Korchemsky and A.~N.~Manashov,
Nucl. \ Phys. \ B {\bf 617} (2001) 375,  arXiv:hep-th/0107193.





\bibitem{Braun:1998id}
V.~M.~Braun, S.~E.~Derkachov and A.~N.~Manashov,
Phys.\ Rev.\ Lett.\  {\bf 81} (1998) 2020
[arXiv:hep-ph/9805225];
V.~M.~Braun, S.~E.~Derkachov, G.~P.~Korchemsky and A.~N.~Manashov,
Nucl.\ Phys.\ B {\bf 553} (1999) 355
[arXiv:hep-ph/9902375].




\bibitem{Belitsky:1999qh}
A.~V.~Belitsky,
Phys.\ Lett.\ B {\bf 453} (1999) 59
[arXiv:hep-ph/9902361];
Nucl.\ Phys.\ B {\bf 558} (1999) 259
[arXiv:hep-ph/9903512];
Nucl.\ Phys.\ B {\bf 574} (2000) 407
[arXiv:hep-ph/9907420].




\bibitem{Beisert:2003yb}
N.~Beisert and M.~Staudacher,
Nucl.\ Phys.\ B {\bf 670} (2003) 439
[arXiv:hep-th/0307042].



\bibitem{Dolan:2003uh}
L.~Dolan, C.~R.~Nappi and E.~Witten,
arXiv:hep-th/0308089.





\bibitem{Derkachov:2000ne}
S.~Derkachov, D.~Karakhanian and R.~Kirschner,
Nucl.\ Phys.\ B {\bf 583} (2000) 691
[arXiv:nlin.si/0003029].

\bibitem{Derkachov:2001sx}
S.~E.~Derkachov, D.~Karakhanyan and R.~Kirschner,
Nucl.\ Phys.\ B {\bf  618} (2001) 589,\
arXiv:nlin.si/0102024.




\bibitem{Karakhanyan:2001wr}
D.~Karakhanyan, R.~Kirschner and M.~Mirumyan,
Nucl.\ Phys.\ B {\bf 636} (2002) 529
[arXiv:nlin.si/0111032].










\bibitem{KS80} P.P. Kulish and E.K. Sklyanin, Zap. Nauchn. Semin. LOMI 95
(1980) 129.

\bibitem{TTF} V.O. Tarasov, L.A. Takhtadjian and L.D. Faddeev, Theor. Math.
Phys. 57 (1983) 163-181.

\bibitem{Nankai} E.K. Sklyanin, "Quantum Inverse Scattering Method", in
{\sl Quantum Groups and Quantum Integrable Systems}, (Nankai lectures),
ed. Mo-Lin Ge, pp. 63-97,
World Scientific Publ., Singapore 1992, [hep-th/9211111]

\bibitem{LesHouches} L.D. Faddeev, Les Houches lectures 1995,
hep-th/9605187.










\bibitem{Manin}
E.E. Demidov, Yu.I. Manin, E.E. Mukhin and D.V. Zhdanovich, Progr.
Theor. Phys. Suppl. {\bf 102} (1990) 203.

\bibitem{Zakrewski}
S. Zakrewski, Lett. Math. Phys. {\bf 22 } (1991) 287.

\bibitem{Stolin}
A. Stolin, Math. Scand. {\bf 69} (1991) 56.

\bibitem{Wess}
H. Ewen, O.V. Ogievetsky and J. Wess, Lett. Math. Phys. {\bf 22} (1991)
297;
H. Ewen and O. Ogievetsky,  {\sl Jordanian solutions of simplex equations},
hep-th/9211026



\bibitem{Kupershmidt} B.A. Kupershmidt, J Phys A Math Gen {\bf 25}
(1992) L1239-L1244.

\bibitem{Ohn} Ch. Ohn, Lett. Math. Phys. {\bf 25} (1992) 85.


\bibitem{Gerstenhaber}
M. Gerstenhaber, A. Giaquinto, S.D. Schack,
in Quantum Groups, Proceedings In EIMI, P. Kulish (ed.), Lecture Notes
in Math., No.1510, p. 9, Springer 1992.

\bibitem{Ogievetsky} O.V. Ogievetsky, Suppl. Rendiconti Cir. Math. Palermo, 
Serie II {\bf37} (1993) 185.








\bibitem{DrinfeldT} V.G. Drinfeld,
Soviet Math. Dokl. {\bf 27} (1983) 68.
\bibitem{DrinfeldY}
V.G. Drinfeld, Soviet Math. Dokl. {\bf 32} (1985) 254 and
                      {\it ibid.} {\bf 36} (1985) 212






\bibitem{KS} P.P. Kulish and A. A Stolin,
Czech. J. Phys. {\bf 47} (1997) 123: and 1207.
q-alg/9608011; and
 q-alg/9708024.

\bibitem{KhST}
S. Khoroshkin, A. Stolin, V. Tolstoy, in:{\sl From Field Theory to Quantum
Groups}, eds. B. Jancewicz and J. Sobczyk, World Scientific (1996), pp.
53-77. q-alg/9511005.




\bibitem{Abdesselam}
B. Abdesselam, A. Chakrabarti, R. Chakrabarti, Mod. Phys. Lett. A {\bf
11} (1996) 2883.



\bibitem{Jeugt} J. Van der Jeugt,J. Phys. A: Math. Gen. {\bf 31}
(1998) 1495-1508, q-alg/9703011; 
Czech. J. Phys. {\bf 47} (1997) 1283-1289,
 q-alg/9709005.



\bibitem{Vladimirov}
A.A. Vladimirov, Mod. Phys. Lett. A {\bf 8} (1993) 2573.

\bibitem{Ballesteros}
A. Ballesteros, F.J. Herranz, M.A. del Olmo, M. Santander,
Journ. Math. Phys. {\bf 35} (1994) 4928; and
Phys. Lett B {\bf 351} (1995) 137.



\bibitem{Lukierski}
J. Lukierski, P. Minnaret and M. Morzrzymas, Phys. Lett. B {\bf 371}
(1996) 215,
q-alg/9507005; \\
J.~Lukierski,
`From noncommutative space-time to quantum relativistic symmetries with
fundamental mass parameter,''
arXiv:hep-th/0112252; \\
J.~Lukierski, V.~Lyakhovsky and M.~Mozrzymas,
Phys.\ Lett.\ B {\bf 538} (2002) 375
[arXiv:hep-th/0203182].



\bibitem{BD}
A.A. Belavin and V.G. Drinfeld,
Funct. Anal. Appl. {\bf 16} (1982) No.3, 1-29.




\end{thebibliography}
\end{document}